\documentclass[12pt]{iopart}
\pdfoutput=1
\usepackage{color}
\usepackage{graphicx}
\newcommand{\revise}[1]{{\color{black} {#1}}}

\begin{document}

\title[]{Criticality and correlated dynamics at the
irreversibility transition in
periodically driven colloidal suspensions}

\author{Elsen Tjhung \& Ludovic Berthier}

\address{Laboratoire Charles Coulomb, UMR 5221, CNRS and 
Universit\'e Montpellier, Montpellier, France}

\begin{abstract}
One possible framework to interpret the irreversibility transition observed in periodically driven colloidal suspensions is that of 
a non-equilibrium phase transition towards an absorbing reversible state at low amplitude of the driving force.
We consider a simple numerical model for driven suspensions which 
allows us to characterize in great detail a large body of physical observables that 
can be experimentally determined to assess the existence and universality class of such a non-equilibrium phase transition. 
Characterizing the behaviour of static and dynamic correlation functions both in real and Fourier space 
we determine in particular several critical exponents for our model, which take values 
that are in {\color{black} good} agreement with the universality class of directed percolation. 
We also provide a detailed analysis of single-particle and collective dynamics of the system near the phase transition, 
which appear intermittent and spatially correlated over diverging timescales and lengthscales, 
and provide clear signatures of the underlying criticality. 
\end{abstract}

\section{Introduction}

Non-equilibrium phase transitions have been gaining interests in recent years \cite{hinrichsen,lubeck-review,lubeck-book}. 
Whereas theoretical models can be organised in well-studied universality classes, 
the interpretation of experimental realisations is always more challenging, 
due to the inherent complexity of an experimental set-up compared to the simplicity of theoretical models. 
In this respect, the experimental study by Pine and coworkers \cite{Pine-Nat,Pine-Nat-Phys,CortePRL} of a relatively simple driven suspension of non-Brownian particles displaying an apparent continuous phase transition towards an absorbing state offered a very promising new path for studying experimentally non-equilibrium phase transitions~\cite{bartolo-nat-comms,ganapathy,bartolo-hyperuniform,filippidi,gollub}. 
In the original work~\cite{Pine-Nat}, the authors studied a colloidal suspension driven out of equilibrium by shearing the system periodically with some frequency $\omega_{0}$ and a maximum shear amplitude $a$. It has been found that below some critical shear amplitude $a_{c}$ (which depends on the volume fraction $\phi$), 
the system evolves into an absorbing state in which all particles move reversibly in response to the external periodic shear.
In other words, when observed stroboscopically, single-particle motion appears to be frozen and thus the diffusion constant $D$ is zero. 
On the other hand above the critical shear amplitude $a_{c}$, the particle motion becomes irreversible and appears diffusive at long times from stroboscopic images. 
This qualitative transformation  was suggested to correspond to an nonequilibrium phase transition from 
an absorbing reversible state with $D=0$ to an irreversible diffusive 
phase with $D>0$, and the results were compatible with a continuous phase transition~\cite{Pine-Nat,Pine-Nat-Phys}.  
Interestingly at high density (not considered in this paper), 
such transitions have also been suggested to be associated with plastic yielding \cite{ganapathy,cipelletti,takeshi,keim}.

Crucially, these colloidal particles are suspended in a very viscous solvent such that thermal effect is negligible over the duration of the experiment. 
The origin of irreversibility was, in fact, attributed to collisions between the particles provoked by the  shear flow~\cite{Pine-Nat-Phys}. 
From this interpretation, Cort\'e {\it et al.} then constructed a simple numerical model for their experiment on sheared non-Brownian suspensions, where microscopic irreversibility upon collisions
in the real system is converted into a stochastic rule for binary collisions in the theoretical model~\cite{Pine-Nat-Phys}.
(Long range hydrodynamic interactions were shown to play little role \cite{butler1,butler2}.)
However, it has also been suggested by several authors 
\cite{bartolo-nat-comms,sundararajan,bartolo-kurchan,reichhardt3,o-hern} that 
the onset of irreversibility could be due to chaotic motion of the particles rather than a nonequilibrium phase transition. 
In other words, macroscopic irreversibility could arise when the Lyapunov exponent is larger than some threshold  without any diverging lengthscales or timescales. 
Therefore further experimental studies are necessary to show if there is, in fact, 
any diverging lengthscale in such systems. 
While both scenario may coexist in the original experiment, 
we have recently suggested a number of possible experimental measurements 
that should be able to determine the main source of the irreversibility transition observed experimentally~\cite{elsen1}. \revise{We give here an extended 
account of these numerical findings.}

The experimental suggestions we made were based on the demonstration 
that a simpler version of the model by Cort\'e {\it et al.}
displays a large number of experimentally accessible signatures
of the criticality associated to the phase transition to an absorbing state. 
In the present paper, we provide a very extensive numerical study 
of this simplified model, which differs from the original one
by the fact that stochastic rules for binary collisions are made 
isotropic~\cite{elsen1,schmiedeberg}. This small 
change in the definition allows us to characterise more accurately the 
critical properties of the system, which are not yet fully resolved.
It has been initially 
suggested by Menon and Ramaswamy \cite{ramaswamy} that
such periodically driven systems may belong to the universality class
of conserved directed percolation (CDP). The conserved quantity, in
this case, refers to the total number of particles. However up to
date, there has been no accurate measurements of the critical exponents
which may suggest a CDP universality class. For instance, the critical
exponent for the order parameter $\beta$ is defined to be: 
$D\sim(a-a_{c})^{\beta}$ as $a\rightarrow a_{c}^{+}$. 
Different values of $\beta$ have been reported in both experimental and numerical studies of  bidimensional periodically sheared suspensions: 
$\beta=0.45\pm0.02$ \cite{Pine-Nat-Phys}, 
$\beta=0.67\pm0.09$ \cite{ganapathy},
$\beta=0.75\pm0.02$ \cite{schmiedeberg}, 
$\beta=0.59$ \cite{reichhardt}, 
$\beta=0.6\pm0.06$ \cite{reichhardt2} and
$\beta=0.69\pm0.11$ \cite{fiocco,fiocco-foffi}. 
However from these values of $\beta$, we are still unable to determine if
the system belongs to the conserved directed percolation (CDP) or
directed percolation (DP) universality class. The value of $\beta$
in DP is $0.58$ and in CDP is $0.64$ for 2D systems \cite{lubeck-review}.
\revise{The first aim of the present paper} 
is to accurately measure $\beta$ as well as
other critical exponents discussed below, to more precisely characterize
the universality class of the model.

Another frequently measured quantity is a timescale $\tau$ which
diverges close to criticality with a power law: 
$\tau\sim(a-a_{c})^{-\nu_{\parallel}}$.
The timescale $\tau$, for instance, 
can be defined as the timescale for the system
to reach a steady state~\cite{Pine-Nat-Phys,ganapathy,reichhardt2}.
This timescale has been measured in the literature giving varying values
of the critical exponents: 
$\nu_{\parallel}=1.33\pm0.02$ \cite{Pine-Nat-Phys},
$\nu_{\parallel}=1.1\pm0.3$ \cite{ganapathy}, 
$\nu_{\parallel}=1.30\pm0.06$ \cite{schmiedeberg}, and 
$\nu_{\parallel}\simeq1.36\pm0.06$ \cite{reichhardt2}
(all are in 2D). In this paper, we shall introduce four other ways
of measuring $\tau$ independently. In particular, we will demonstrate
that the results of these measurements are consistent to each other,
giving the same critical exponent $\nu_{\parallel}$. Another 
important critical exponent is associated with the diverging lengthscale: 
$\xi\sim(a-a_{c})^{-\nu_{\perp}}$.
However, $\nu_{\perp}$ has not yet been measured in literature. In
this paper, both static and dynamic correlation lengths $\xi$ are
carefully studied for the first time and characterised quantitatively.
In particular, they are both shown
to diverge close to criticality with the same exponent $\nu_{\perp}$.
A related critical exponent is the one of the susceptibility 
$\chi$ which quantifies the variance of the global fluctuations of 
the order parameter, and obyes $\chi \sim (a- a_c)^{-\gamma}$. 
Again, the exponent $\gamma$ has, to our knowledge, not been measured
before, \revise{because it demands a proper (i.e. grand-canonical) 
estimate of the global fluctuations of the order parameter.}
\revise{The second aim of this work is to show that 
a large number of observables exist that can reveal directly the 
criticality of the model.}

Thus both our model and the one introduced by Cort\'e \emph{et al.}~\cite{Pine-Nat-Phys}
show a true dynamical phase transition from an irreversible to a reversible phase accompanied by diverging timescales, lengthscales and global fluctuations. 
By simultaneously measuring all critical exponents using various methods that appear to be internally consistent, 
we find {\color{black} these values to be consistent with both directed percolation (DP) and conserved directed percolation (CDP/Manna) universality class after 
appropriate finite size scaling. We emphasize that both universality classes 
share critical exponents that are very close to one another and are hard to distinguish numerically. Our results seem to favor the DP universality class, despite the presence of particle conservation, which is generally viewed as the key ingredient responsible for the emergence of the CDP universality 
class~\cite{lubeck-review,lubeck-book}. We discuss this finding 
more extensively in the conclusion of the article.}

Additionally, our simplified model also shows a true 
hyperuniformity at criticality,
consistent to the conventional shear-based 
Cort\'e model~\cite{bartolo-hyperuniform,levine,frenkel-hyperuniform},
which we show by a diverging hyperuniform lengthscale 
approaching criticality, as determined experimentally very 
recently~\cite{bartolo-hyperuniform}. We find, however, that this
hyperuniform lengthscale is apparently 
unrelated to other critical lengthscales in the model.
\revise{It is therefore unclear to us whether a deep connection exists between
critical and hyperuniform fluctuations, although it does suggest a 
qualitative connection between fluctuations of the density and 
that of the order parameter.}

Finally, we shall also point out a similarity between 
the dynamic criticality displayed in periodically
driven systems near an irreversibility transition 
and dense liquids near a glass transition \cite{berthier-book,RMP}. We 
will show that both types systems
exhibit a very heterogenous dynamics in space and time (which becomes
critical at the reversible-irreversible transition).
We find in particular that
single-particle dynamics becomes intermittent and strongly non-Fickian, 
and that collective dynamics
becomes spatially correlated over diverging lengthscales.
The analogy between the two
types of systems suggests that particle-based measurements
and observables developed for glassy materials could
prove useful in driven suspensions. These tools could
in particular reveal whether the `singularity-free' 
explanation based on the Lyapunov instability is experimentally
relevant, as no special critical feature should appear in this case.

Our paper is organised as follows. 
In Sec.~\ref{model} we define the numerical model used in this study.
The phase diagram and the order parameter critical exponent 
are analysed in Sec.~\ref{sec:order-parameter}.
Static lengthscales and global fluctuations are measured 
in Sec.~\ref{sec:static}.
We study single-particle and collective dynamics 
in Sec.~\ref{sec:dynamic} and conclude in Sec.~\ref{sec:conclusion}.

\section{Model}
\label{model}

\begin{figure}
\begin{centering}
\includegraphics[scale=0.6]{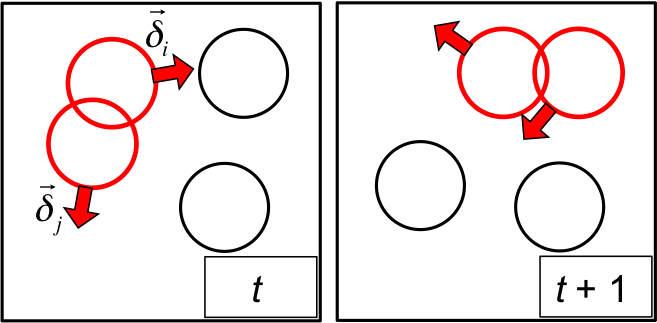}
\par\end{centering}
\caption{The model considered in our work: 
Particles overlapping at time $t$ (red) are labelled `active' and are each given an independent random displacement $\vec \delta_i$.
Particles with no overlap (black) are labelled `passive' and are 
immobile, but may become active at a later time.} 
\label{fig:model}
\end{figure}

The model considered in this paper is shown schematically in Fig.~\ref{fig:model}.
We consider a 2-dimensional assembly of spherical particles of diameter $\sigma$ in a box of linear size $L$ with periodic boundary conditions. 
The system is initiated from a completely random configuration, where particle overlaps may exist.
At each time step $t$, particles which overlap with at least one neighbour are labelled `active' and
we simultaneously move each active particle $i$ by an independent random displacement $\vec \delta_i$:
\begin{equation}
\vec \delta_i = \epsilon_i \hat e_i, 
\label{eq:delta}
\end{equation}
where $\hat e_i$ is a unit vector whose orientation is distributed randomly on a unit circle and 
the magnitude $\epsilon_i$ is a random number distributed uniformly on the interval $[0,\delta]$. 
Particles which do not overlap, on the other hand, are labelled `passive' and remain immobile though 
they may become active and mobile later.
The time $t$ is then incremented by one unit.
The model has two control parameters:
the area fraction $\phi=\pi N \sigma^2 / 4L^2$ and
the maximal amplitude of the random kicks $\delta$.
We use $\sigma$ as the unit length and we vary the area fraction by changing the number of particles $N$
while keeping the system size fixed at $L=280$ (unless mentioned otherwise).
For this system size, we do not find significant 
finite size effect for the range of parameters that we use below.
The value of the critical density obviously depends on 
the magnitude of the random kick $\delta$. We shall 
establish the phase diagram $\phi_c(\delta)$ in 
Sec.~\ref{sec:order-parameter}, but will conduct most 
of the our numerical studies of the critical behaviour of the system
using $\delta=0.5$, for which $\phi_{c} \simeq 0.375$.
We find that critical exponents are insensitive to the 
specific choice of $\delta$ and this value thus represents 
a compromise between too small 
values of $\delta$ which increase the computational time
(because particles move very little),  
and much larger values of $\delta$, which become unphysical 
when $\delta  \gg  \sigma$. 

Our model represents a simplified isotropic version of the periodically sheared suspension model considered in Ref.~\cite{Pine-Nat-Phys}.
In \cite{Pine-Nat-Phys}, random kicks are given to all particles which collide during a shear deformation cycle.
This original rule is in fact equivalent to giving a random kick to each particle having at least one neighbour in an anisotropic area near its centre~\cite{schmiedeberg}.
In our model, we consider that this area is circular with $\sigma$ representing its diameter.
This small simplification makes the determination of the critical properties of the model simpler because it prevents the development of locally anisotropic correlations~\cite{schall}, 
which could affect the numerical value of the measured exponents, but not the overall qualitative behaviour that we report.
Our model may also represent an experimental situation in which a non-Brownian colloidal suspension is driven periodically by a periodic change of the particle diameters, leading to irreversible collisions.
Such experiments could be realised using thermo-sensitive colloidal particles~\cite{ganapathy,percier,reviewPNIPAM}.

{\color{black}
We perform each simulation run over $1.6 \times 10^6$ timesteps. 
Averages are taken over $10^6$ timesteps in steady state.
Close to criticality, we perform $4$ independent simulation runs and if at least $1$ run reaches an absorbing state before the maximum timestep is reached,
the data for that density are discarded.
The measurement itself is taken from a single run, using an extensive time averaging procedure.} 

\section{Order parameter and phase diagram 
\label{sec:order-parameter}}

We define the order parameter as the fraction of active particles at time $t$: 
\begin{equation}
f_{a}(t) = \frac{N_{a}(t)}{N}, \label{eq:fa}
\end{equation}
where $N_{a}(t)$ is the total number of active/mobile particles (see Fig.~\ref{fig:model}). 

\begin{figure} 
\begin{centering}
\includegraphics[width=1\columnwidth]{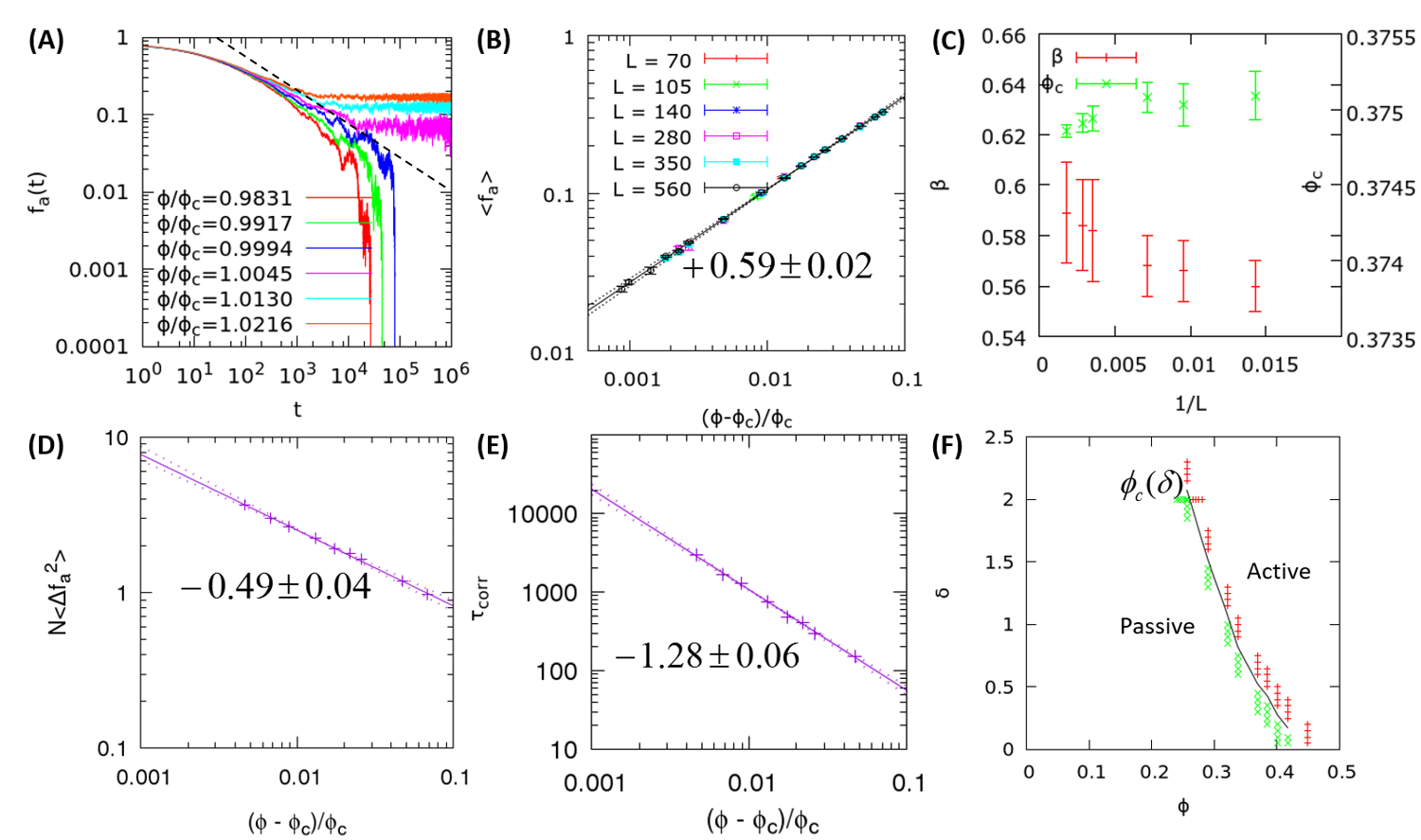}
\par\end{centering}
\caption{
\textbf{(A)}
Time dependence of the fraction of active particles 
from a random initial configuration.
In the long-time limit, $f_a$ vanishes for $\phi<\phi_{c}$,
fluctuates around a well-defined average above $\phi_c$. 
At criticality, a power law is observed, $f_a(t)\sim t^{-\alpha}$ 
with the exponent $\alpha\simeq0.45$ 
{\color{black} (dashed line)}.
\textbf{(B)}
Algebraic vanishing of 
the average order parameter $\langle f_{a}\rangle$ with density {\color{black} for different system sizes $L$},
$\langle f_{a}\rangle \sim(\phi-\phi_{c})^{\beta}$, with $\beta\simeq0.59\pm0.02$. 
{\color{black} The dashed lines indicate uncertainty in the critical exponent.
\textbf{(C)}
System size dependence of the critical exponent $\beta$ and 
critical density $\phi_c$.}
\textbf{(D)}
Diverging fluctuations of the order parameter at steady state,
$N\langle\Delta f_a^2\rangle\sim(\phi-\phi_{c})^{-\gamma}$ 
with $\gamma \simeq 0.50$. 
\textbf{(E)} 
Diverging correlation time from steady state fluctuations 
of the order parameter, 
$\tau_{corr}\sim(\phi-\phi_{c})^{-\nu_\parallel}$ 
with $\nu_\parallel\simeq1.27$.
\textbf{(F)}
Phase diagram on the parameter space $(\phi,\delta)$ 
separating the active phase (pluses) where the number of active 
particles remain finite at steady state and 
the absorbing phase (crosses) where the number of active particles 
drops to zero at steady state. The full line is a continuous line 
of equivalent second-order phase transitions.}
\label{fig:fa}
\end{figure}

Fig.~\ref{fig:fa}(A) shows the evolution of the order parameter $f_a(t)$ (from an initially random configuration at $t=0$). 
Below some critical density $\phi<\phi_{c}$, 
the number of active/mobile particles goes to zero as 
$t\rightarrow\infty$ (passive phase). 
The passive phase is also an absorbing state. 
This means once the number of active/mobile particles goes to zero, 
it is not possible to recover mobility. 
Physically, all overlaps (which are the source of mobility in our system) 
have been removed. 
On the other hand above the critical density $\phi>\phi_{c}$, 
the number of active/mobile particles remains finite and fluctuates around its mean $\langle f_{a}\rangle $
in the steady state (active phase). 
At criticality $\phi=\phi_c$,
the order parameter decays as power law: $f_a(t)\sim t^{-\alpha}$ with exponent $\alpha\simeq0.45$, \revise{as indicated in Fig.~\ref{fig:fa}(A).}
This value of $\alpha$ is close to the one reported in 2D directed percolation and conserved directed percolation universality classes~\cite{lubeck-review}. 

The mean value $\langle f_{a}\rangle$ is simply defined to be the time average at steady state:
\begin{eqnarray}
\langle f_{a}\rangle & = & \frac{1}{T}\int_{t''}^{t''+T}dt'\, f_{a}(t'),      \label{eq:fa_average}
\end{eqnarray}
where $t''$ is larger than the time it takes for the system reaches a steady state.
Fig.~\ref{fig:fa}(B) shows the average order parameter $\langle f_{a}\rangle$ as a function of packing fraction $\phi$
{\color{black} at different system sizes $L$.
The $y$-errorbar in $\langle f_a \rangle$ is taken from square root of the fluctuation squared in $f_a(t)$ divided by the number of independent measurements
(which is $T$ divided by the correlation time).}
The average order parameter is zero in the passive phase ($\phi<\phi_{c}$) and
positive in the active phase ($\phi>\phi_{c}$). 
Furthermore from Fig.~\ref{fig:fa}(B), 
we find that the average order parameter decreases continuously to zero at critical density $\phi_c$ with power law behaviour: 
\begin{equation}
\langle f_{a}\rangle \sim(\phi-\phi_{c})^{\beta},\quad {\rm as}\,\,\,\phi\rightarrow\phi_{c}^{+}.
\end{equation}
{\color{black} The critical exponent $\beta$ is measured to be  $\simeq0.59\pm0.02$ and the critical density is $\phi_c\simeq0.375\pm0.001$.
Both are found by fitting a best straight fit line on the log-log scale (see Fig.~\ref{fig:fa}(B)). 
The dashed lines in Fig.~\ref{fig:fa}(B) indicate the $\pm0.02$ uncertainty in the slopes or the critical exponent $\beta$.
The system size dependence of $\beta$ and $\phi_c$ are plotted in Fig.~\ref{fig:fa}(C).
We found no significant finite size effects for these measurements, which depend only weakly on the system size. 
From the finite size scaling, we find that the critical exponent $\beta$ is consistent to both 2D directed percolation universality class ($\beta\simeq0.5834$)
and conserved directed percolation or Manna universality class ($\beta\simeq0.639$)~\cite{lubeck-review}, although our data do appear to lie closer to the DP value. Although we cover a significant range of system sizes and correlation lengthscales (see below), we can of course not exclude the existence of a crossover at even larger lengthscale towards the CDP value, and even larger-scale simulations would be needed to see this. This would represent, we believe, a very significant numerical effort.}

As we can see from Fig.~\ref{fig:fa}(A), the fraction of active particles fluctuates around its mean value $\langle f_{a}\rangle$ in the active phase at steady state. 
This fluctuation becomes larger and diverges as we approach the critical density from above $\phi\rightarrow\phi_{c}^{+}$.
We define the average fluctuation squared (or variance of the order parameter) to be:
\begin{equation}
N\left\langle \Delta f_a^2\right\rangle =N\left\langle (f_{a}(t')-\left\langle f_{a}\right\rangle )^{2}\right\rangle, \label{eq:Dfa_average}
\end{equation}
where the angle bracket indicates time averaging over $t'$ at steady state, 
similar to Eq.~(\ref{eq:fa_average}). 
We plot the average fluctuation squared, $N\left\langle \Delta f_a^2\right\rangle $, as a function of $\phi$ in Fig.~\ref{fig:fa}(D) which 
again exhibits a power law behaviour close to the critical point: 
\begin{equation}
N\left \langle \Delta f_a^2\right\rangle \sim(\phi-\phi_{c})^{-\gamma}, \quad 
{\rm as}\,\,\,  \phi\rightarrow\phi_{c}.
\end{equation} 
{\color{black} Similarly, the dashed lines indicate uncertainty in the critical exponent.}
However in this case, the measured 
critical exponent ($\gamma\simeq0.49$) appears 
quite different from that reported \revise{in both DP and CDP} universality 
classes ($\gamma=0.30$ and 0.37 respectively~\cite{lubeck-review}).
\revise{This is because the global fluctuations of the order parameter 
in Fig.~\ref{fig:fa}(D) are measured in the constrained ensemble where the 
total number of particles $N$ is fixed, which obviously
affects the global fluctuations of any physical quantity~\cite{lebo}.}
In Sec.~\ref{sub:grand-canonical}, we shall look at a more general fluctuation of the order parameter, computed in the grand-canonical ensemble. 
In that case, we will find that the critical exponent becomes very close 
to that of 2D directed percolation. 

The time auto-correlation function of the order parameter is defined as:
\begin{equation}
C(t)=\frac{\langle f_{a}(t'+t)f_{a}(t')\rangle - \langle f_{a}\rangle\langle f_{a}\rangle }{\left\langle \Delta f_{a}^2\right\rangle }, \label{eq:C_ff}
\end{equation}
where the angle average indicates time averaging over $t'$ at steady state as before. 
At $t=0$, the value of $C(t)$ is equal to $1$ (maximum correlation) and,
as $t$ increases, $C(t)$ decreases and eventually goes to zero in the limit $t\rightarrow\infty$ (completely decorrelated).
The auto-correlation time of the order parameter $\tau_{corr}$ is defined by setting $C(\tau_{corr})=e^{-1}$. 
The correlation time is shown to diverge near the critical point $\phi_{c}$ with a power law: 
\begin{equation}
\tau_{corr}\sim(\phi-\phi_{c})^{-\nu_{\parallel}},
\end{equation} 
where $\nu_{\parallel}=1.27$ (see Fig.~\ref{fig:fa}(E)). 
Again this value of $\nu_\parallel$ is close to that of 2D directed percolation.
Note that we report below the behaviour of 
other timescales and all these timescales were found 
to diverge with the same critical exponent $\nu_\parallel$, which 
we numerically find is close to $\nu_\parallel\simeq1.26$. 
Such an agreement between critical exponents 
also holds for lengthscales (except for hyperuniform length), and similarly, 
we shall label the critical exponent associated with 
lengthscales with the unique notation $\nu_\perp$.

Fig.~\ref{fig:fa}(F) shows the phase diagram on the parameter space $(\phi,\delta)$ showing the critical line $\phi_{c}(\delta)$
which separates the active phase from the passive phase. 
Each point on the line is a continuous/second order transition from passive to active phase. 
Furthermore, we have also found similar critical exponent $\beta\simeq0.57\pm0.02$ for the order parameter
$\left\langle f_{a}\right\rangle (\phi,\delta)$ 
at different points along the critical line going horizontally along $\phi$ or vertically along $\delta$. 
For the rest of this paper, we shall fix $\delta=0.5$ for which the critical density is $\phi_{c}\simeq0.375$.

\section{Static properties}

\label{sec:static}

\subsection{Structure factors and correlation lengthscales}

\label{sec:structure-factors}

\begin{figure} 
\begin{centering}
\includegraphics[scale=0.6]{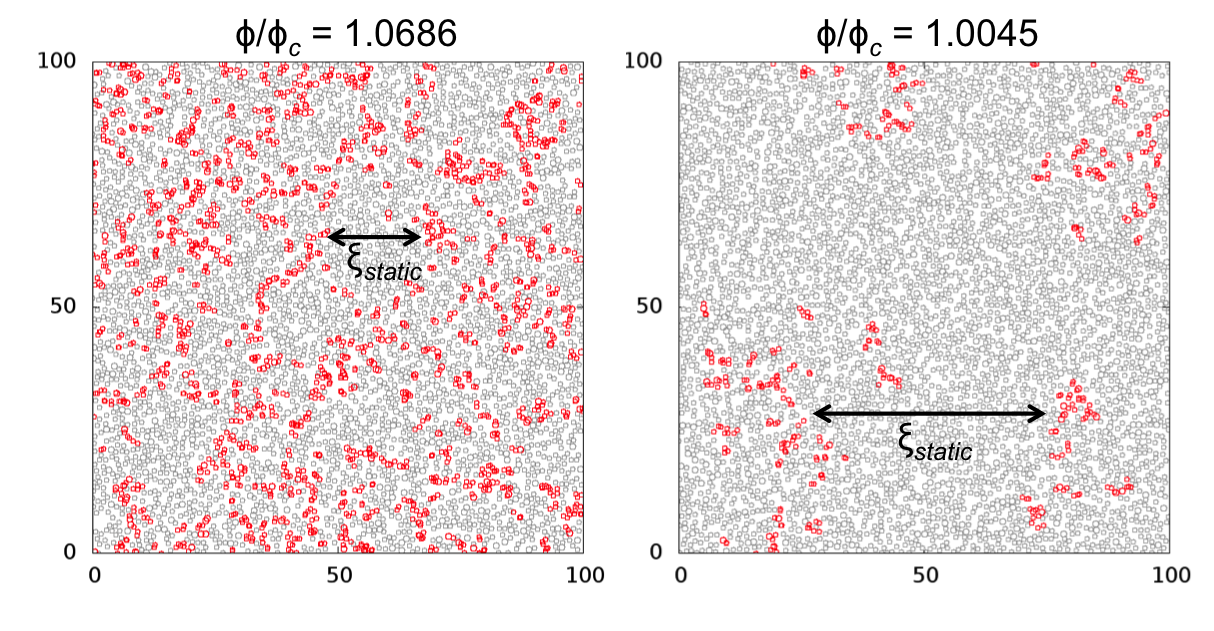}
\par\end{centering}
\caption{
Snapshots of the active phase far above the critical density (left)
and closer to the critical density (right). 
The red circles indicate particles which are active. 
As can be seen from the snapshots, we observe a growing static lengthscale as 
we approach the critical density of $\phi_{c}\simeq0.375$.} 
\label{fig:snapshot-static}
\end{figure}

Fig.~\ref{fig:snapshot-static} shows the snapshot of the active phase far from critical density (left panel) 
and close to critical density (right panel), taken at steady state. 
The red circles indicate active/mobile particles and the grey circles indicate passive/immobile particles. 
Note that active particles can become passive after some time and \emph{vice versa}.
As we can see from the figure, active particles tend to cluster together and 
we shall define a `static' lengthscale to be the typical size of these clusters, or equivalently, typical distance between two clusters. 
Comparing the two panels in Fig.~\ref{fig:snapshot-static}, we observe a growing static lengthscale
as we approach the critical density from above ($\phi\rightarrow\phi_{c}^{+}$).
In this section, we shall discuss how these static lengthscales can be measured.
Note that in the experimental situation, the activity of the particle at time $t$ is determined over one cycle of oscillation.
Thus the lengthscales discussed in this section are not purely static but instead determined by the dynamics of the particles over one cycle of oscillation. 
However for the purpose of the discussion, we shall call the lengthscales in this section as almost `static' quantities.
This is to distinguish such lengthscales from dynamic lengthscales in Sec.~\ref{sec:collective-dynamics}, 
which are measured over much larger time delays comprising many cycles.

Suppose we denote the number density of the active and passive particles to be $n_{a}(\mathbf{r},t)$ and $n_{p}(\mathbf{r},t)$ respectively.
These can be written explicitly as:
\begin{eqnarray}
n_{a}(\mathbf{r},t) & = & \sum_{i=1}^{N}\delta(\mathbf{r}-\mathbf{r}_{i})f_{i}(t)=\sum_{i=1}^{N_{a}}\delta(\mathbf{r}-\mathbf{r}_{i}^{a}(t)), \nonumber \\
n_{p}(\mathbf{r},t) & = & \sum_{i=1}^{N}\delta(\mathbf{r}-\mathbf{r}_{i})(1-f_{i}(t))=\sum_{i=1}^{N_{p}}\delta(\mathbf{r}-\mathbf{r}_{i}^{p}(t)), \label{eq:density}
\end{eqnarray}
where $f_{i}(t)=1$ if particle $i$ is active at time $t$ and $f_{i}(t)=0$ if particle $i$ is passive at time $t$; 
$\mathbf{r}_{i}^{a}(t)$ and $\mathbf{r}_{i}^{p}(t)$ are the positions (at time $t$) of the active and passive particles respectively. 
The average densities are simply:
\begin{equation}
\left\langle n_{\alpha}(\mathbf{r},t)\right\rangle = \frac{\left\langle N_{\alpha}\right\rangle }{V},  \label{eq:average-density}
\end{equation}
where $\alpha=a,p$ refers to either active or passive particles.
$V$ is the volume of the system and 
$\langle N_{a/p}\rangle$ is the averaged total number of active/passive particles respectively. 
The structure factors between the active particles $S_{aa}(q)$ and between the passive
particles $S_{pp}(q)$ are defined to be \cite{bhatia-thornton}:
\begin{equation}
S_{\alpha\alpha}(q)=\frac{1}{\langle N_{\alpha} \rangle}\left\langle \delta n_{\alpha}^{*}(q)\delta n_{\alpha}(q)\right\rangle, \label{eq:structure-factor}
\end{equation}
where $\delta n_{\alpha}(q)$ is the Fourier transform of $\delta n_{\alpha}(\mathbf{r})=n_{\alpha}(\mathbf{r})-\left\langle n_{\alpha}\right\rangle $.
By substituting Eq.~(\ref{eq:density}) into (\ref{eq:structure-factor}),
the structure factor can be computed explicitly as:
\begin{equation}
S_{\alpha\alpha}(q)=\left\langle \frac{1}{N_{\alpha}}\sum_{i=1}^{N_{\alpha}}\sum_{j=1}^{N_{\alpha}}e^{i\mathbf{q}\cdot(\mathbf{r}_{i}^{\alpha}-\mathbf{r}_{j}^{\alpha})}\right\rangle -\left\langle N_{a}\right\rangle \delta_{\mathbf{q},0}. \label{eq:structure-factor-2}
\end{equation}
The limit $q\rightarrow0$ of the structure factors tells us about
the fluctuations of the number of particles since:
\begin{equation}
\frac{\left\langle \Delta N_{\alpha}^{2}\right\rangle }{\left\langle N_{\alpha}\right\rangle }=\frac{1}{\left\langle N_{\alpha}\right\rangle }\left\langle \delta n_{\alpha}^{*}(q\rightarrow0)\delta n_{\alpha}(q\rightarrow0)\right\rangle =S_{\alpha\alpha}(q\rightarrow0). \label{eq:delta-N-squared}
\end{equation}

Finally we can define the mixed structure factor between the active
and passive particles to be:
\begin{eqnarray}
S_{pa}(q) & = & \frac{1}{\sqrt{\left\langle N_{p}\right\rangle \left\langle N_{a}\right\rangle }}Re\left\langle \delta n_{p}^{*}(q)\delta n_{a}(q)\right\rangle \nonumber \\
 & = & \frac{1}{\sqrt{\left\langle N_{p}\right\rangle \left\langle N_{a}\right\rangle }}Re\left\langle \sum_{i=1}^{N_{p}}\sum_{j=1}^{N_{a}}e^{i\mathbf{q}\cdot(\mathbf{r}_{i}^{p}-\mathbf{r}_{j}^{a})}\right\rangle -\sqrt{\left\langle N_{p}\right\rangle \left\langle N_{a}\right\rangle }\delta_{\mathbf{q},0}. \label{eq:mixed-structure-factor}
\end{eqnarray}
The limit $q\rightarrow0$ of $S_{pa}(q)$ gives us the cross-correlation 
of the number fluctuations:
\begin{equation}
S_{pa}(q\rightarrow0)=\frac{\left\langle \Delta N_{a}\Delta N_{p}\right\rangle }{\sqrt{\left\langle N_{a}\right\rangle \left\langle N_{p}\right\rangle }}. \label{eq:Spa}
\end{equation}

\begin{figure} 
\begin{centering}
\includegraphics[width=1\columnwidth]{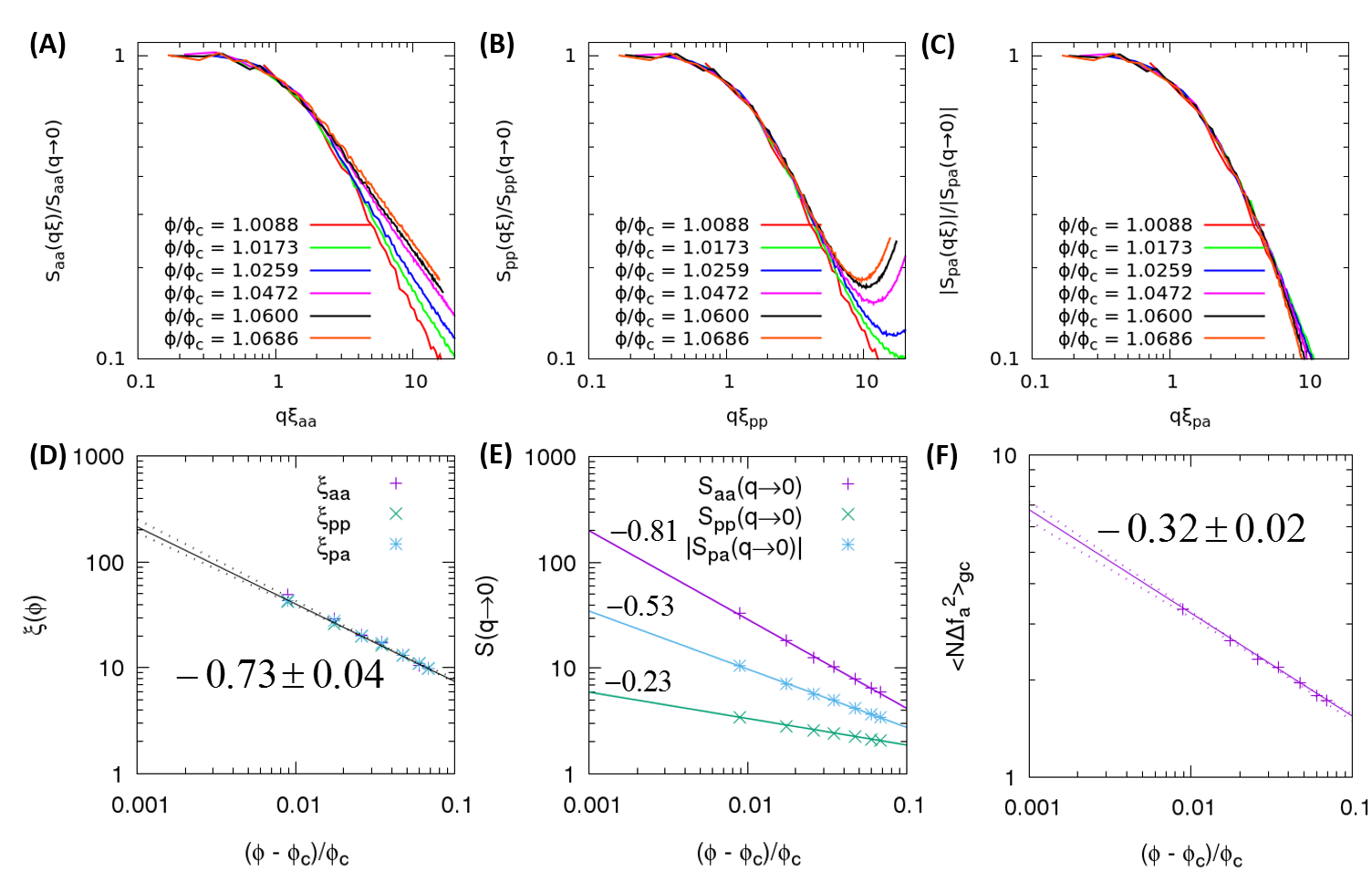}
\par\end{centering}
\caption{
\textbf{(A,B,C)} Scaling analysis of 
partial structure factors between active-active particles $S_{aa}(q)$, 
passive-passive particles $S_{pp}(q)$ and 
passive-active particles $S_{pa}(q)$ using 
Eq.~(\ref{eq:rescaled-structure-factor}). This data collapse
allows for the determination of a correlation
lengthscale $\xi_{\alpha\beta}(\phi)$. 
\textbf{(D)} 
The measured lengthscales $\xi_{\alpha\beta}(\phi)$ diverge as
$\xi \sim(\phi-\phi_c)^{-\nu_\perp}$, with the same critical 
exponent $\nu_\perp\simeq0.73$, close to that of 2D directed percolation. 
{\color{black}
The dashed lines indicate uncertainty in the critical exponent.}
\textbf{(E)}
Global number densitiy fluctuations, $S_{\alpha\beta}(q\rightarrow0)$, 
also diverge with power laws as we approach $\phi_c$, 
with three distinct critical exponents.
\textbf{(F)} The appropriate 
combination of $S_{aa}(q\rightarrow0)$, $S_{pp}(q\rightarrow0)$ 
and $S_{pa}(q\rightarrow0)$ in Eq.~(\ref{eq:fluctuation_fa_2})
provides the order parameter fluctuations, 
$\langle N\Delta f_a^2 \rangle_{gc}$, which  
diverges as  $\chi = \langle 
N\Delta f_a^2 \rangle_{gc}\sim(\phi-\phi_c)^{-\gamma}$ 
with $\gamma\simeq0.31$, in good agreement with 
DP universality class.
\label{fig:structure-factors}}
\end{figure}

In general the fluctuations squared of the number density $S_{\alpha\beta}(q\rightarrow0)$ depend on the packing fraction $\phi$ which diverges as $\phi\rightarrow\phi_{c}^+$.
In fact in critical phenomena, the structure factors can be rescaled by a lengthscale $\xi(\phi)$ such that all structure factors at different packing fractions $\phi$
can be collapsed into a single universal function $F(x)$ which does not depend on $\phi$:
\begin{equation}
\frac{S_{\alpha\beta}(q\xi(\phi))}{S_{\alpha\beta}(q\rightarrow0,\phi)}=F_{\alpha\beta}(q\xi(\phi)), \label{eq:rescaled-structure-factor}
\end{equation}
which should be true as long as $q$ is small compared to $2\pi/\xi$. 
The structure factors of active-active particles $S_{aa}(q)$, passive-passive particles $S_{pp}(q)$ and passive-active particles $S_{pa}(q)$ for different values of $\phi>\phi_{c}$ 
are plotted in Figs.~\ref{fig:structure-factors}(A,B,C). 
As can be seen from the figure, all structure factors from different densities can indeed be collapsed into a single curve, from which, 
a lengthscale $\xi_{\alpha\beta}(\phi$) can be extracted and is shown to diverge with a power law: 
\begin{equation}
\xi\sim(\phi-\phi_c)^{-\nu_\perp}, \quad  {\rm as} \,\,\, 
\phi\rightarrow\phi_c^+,
\end{equation}
(see Fig.~\ref{fig:structure-factors}(D)).
We assume all lengthscales are equivalent: $\xi_{aa}\propto\xi_{pp}\propto\xi_{pa}$
and we found the average critical exponent for the static lengthscale to be $\nu_\perp\simeq0.73$,
consistent to that of 2D directed percolation (see Fig.~\ref{fig:structure-factors}(D)).

Finally, the number density fluctuations of active, passive, or mixed
particles can be found from the limit $S_{\alpha\beta}(q\rightarrow0)$
(see Fig.~\ref{fig:structure-factors}(E)). 
These number fluctuations are also shown to be power law divergent as we approach the critical density with 
three distinct critical exponents as indicated in the figure. 
The ratio of the lengthscales $\xi_{\alpha\beta}$ to number fluctuations $S_{\alpha\beta}(q\rightarrow0)$
is related to another critical exponent which we shall call $\theta_{\alpha\beta}$. 
More precisely, we define $\theta_{\alpha\beta}$ to be:
\begin{equation}
S_{\alpha\beta}(q\rightarrow0)=\xi_{\alpha\beta}^{2-\theta_{\alpha\beta}}.
\label{eq:theta}
\end{equation}
For active-active this value is $\theta_{aa}=0.97$, 
for passive-passive $\theta_{pp}=1.67$ and finally 
for mixed passive-active $\theta_{pa}=1.28$.

\subsection{Grand canonical fluctuations in the order parameter \label{sub:grand-canonical}}

In Sec.~\ref{sec:order-parameter}, we defined the order parameter to be the fraction of active particles at time $t$:
\begin{equation}
f_{a}(t)=\frac{N_{a}(t)}{N}\quad {\rm (canonical)}, \label{eq:fa-canonical}
\end{equation}
where the total number of particles in our system is fixed: $N=constant$ (\emph{i.e.} canonical ensemble). 
Thus, the only source of fluctuations in the order parameter comes from the total number of active particles $N_{a}(t)$. 
Subsequently, we have also found that the canonical fluctuations diverge near the critical density as power law:
$N\langle \Delta f_a^2 \rangle \propto(\phi-\phi_{c})^{-\gamma}$
with a critical exponent $\gamma=0.50$. 

On the other hand, we could have also defined the order parameter in the grand-canonical ensemble where 
the total number of particles is not constant, but fluctuates around its mean value $\langle N \rangle$:
\begin{equation}
f_{a}(t)=\frac{N_{a}(t)}{N(t)}\quad {\rm (grand-canonical)}. 
\label{eq:fa-grand-canonical}
\end{equation}
The mean value of the order parameter $\langle f_{a} \rangle$ is the same whether 
it is taken in the canonical or grand-canonical ensemble: 
$\langle f_{a}\rangle = \langle f_{a} \rangle_{gc}$.
(The subscript $\left\langle \cdot\right\rangle _{gc}$ in this section indicates time averaging in the grand canonical ensemble.) 
However the averaged fluctuations squared of the order parameter are different:
$\langle \Delta f_a^2 \rangle \neq \langle \Delta f_a^2 \rangle_{gc}$
since fluctuations in $N(t)$ will now contribute to fluctuations in $f_{a}(t)$.

We may obtain the grand canonical fluctuation squared of the order parameter from the partial structure factors indirectly as follow.
First we write the order parameter density to be:
\begin{equation}
f_{a}(\mathbf{r},t)=\frac{n_{a}(\mathbf{r},t)}{n_{a}(\mathbf{r},t)+n_{p}(\mathbf{r},t)}. \label{eq:fa-density}
\end{equation}
Thus the average fluctuation over the whole system volume is:
\begin{eqnarray}
\Delta f_{a}(t) & = & \frac{1}{V}\int_{V}dV\,\delta f_{a}(\mathbf{r},t)\nonumber \\
                      & = & \frac{1}{V}\delta f_{a}(q=0,t), \label{eq:delta-fa}
\end{eqnarray}
where $\delta f_{a} = f_{a} - \langle f_{a} \rangle $. 
Therefore the fluctuations squared of the order parameter in the grand-canonical scheme is given by:
\begin{equation}
\left\langle N\Delta f_a^2\right\rangle _{gc}=\frac{\left\langle N\right\rangle }{V^{2}}\left\langle \delta f_{a}^{*}(q\rightarrow0)\delta f_{a}(q\rightarrow0)\right\rangle. \label{eq:fluctuation_fa_1}
\end{equation}
Writing $\delta f_{a}$ in terms of $\delta n_{a}=n_{a}-\left\langle n_{a}\right\rangle $ and $\delta n_{p}=n_{p}-\left\langle n_{p}\right\rangle $, we have:
\begin{equation}
\delta f_{a}=\frac{V}{\left\langle N\right\rangle }\left[(1-\left\langle f_{a}\right\rangle )\delta n_{a}-\left\langle f_{a}\right\rangle \delta n_{p}\right],\label{eq:delta_fa}
\end{equation}
and substituting Eq.~(\ref{eq:delta_fa}) into (\ref{eq:fluctuation_fa_1}), 
the grand canonical fluctuations squared of the order parameter can be expressed in terms of the partial structure factors as follows
\revise{\cite{lebo,bhatia-thornton}:}
\begin{eqnarray}
\left\langle N\Delta f_a^2\right\rangle _{gc} & = & (1-\left\langle f_{a}\right\rangle )^{2}\left\langle f_{a}\right\rangle S_{aa}(q\rightarrow0)+\left\langle f_{a}\right\rangle ^{2}(1-\left\langle f_{a}\right\rangle )S_{pp}(q\rightarrow0) \nonumber \\ 
 && -2\left[\left\langle f_{a}\right\rangle (1-\left\langle f_{a}\right\rangle )\right]^{3/2}S_{pa}(q\rightarrow0). \label{eq:fluctuation_fa_2}
\end{eqnarray}
\revise{Physically, this formula estimates the fluctuations of the order 
parameter that would be measured if we were observing a subsystem of our
model immersed in a much larger system size~\cite{lebo}, as would be done in 
an experiment, for instance. We emphasize that `grand-canonical' does
not mean that the number of particles is no longer conserved, simply 
that the fluctuations inside a subsystem only arise via 
exchange with the reservoir. Therefore, Eq.~(\ref{eq:fluctuation_fa_2})
provides the correct estimate of the global fluctuations
of the order parameter for our model for which the microscopic dynamics
conserves particles.}
Thus, from the limit $q\rightarrow0$ of the partial structure factors $S_{\alpha\beta}$ in Fig.~\ref{fig:structure-factors}(E), 
we obtain the grand canonical fluctuation squared of the order parameter $\langle N\Delta f_a^2 \rangle _{gc}$, shown in Fig.~\ref{fig:structure-factors}(F). 
We now find that the grand-canonical fluctuations squared of the order parameter diverges as $\langle N\Delta f_a^2~\rangle_{gc} \sim 
(\phi-\phi_c)^{-\gamma_{gc}}$ with the critical exponent $\gamma_{gc}\simeq0.32$.
This is in good agreement to that of 2D directed percolation, and 
thus resolves the apparent discrepancy of the canonical critical 
exponent $\gamma \simeq 0.50$ in Sec.~\ref{sec:order-parameter}.

\subsection{Hyperuniform scaling}

Consider a grand-canonical system with volume $V=L^{d}$, where 
$d$ is the dimension of the system ($d=2$ in our case). 
The system can exhange particles with the reservoir and consequently 
the total number of particles fluctuates around its mean $\left\langle N\right\rangle $.
For a Poisson process, the variance of the total number of particles scales as the system volume:
\begin{equation}
\left\langle \Delta N^{2}\right\rangle \sim L^{d}, \label{eq:N-fluctuation}
\end{equation}
whereas for a hyperuniform system, the fluctuation is suppressed:
\begin{equation}
\left\langle \Delta N^{2}\right\rangle \sim L^{d-\lambda}, 
\label{eq:N-fluctuation-hyperuniform}
\end{equation}
where $\lambda\in[0,1]$ is the hyperuniform exponent \cite{stillinger-review}.
The exponent value of $\lambda=1$ has been reported in hard sphere
systems close to the 
jamming transition~\cite{stillinger-prl,berthier-hyperuniform}.
However, it has also been suggested recently in~\cite{ikeda,olsson-teitel} that 
true hyperuniformity might not survive in the limit of large system size at the 
jamming transition.
On the other hand in periodically sheared suspension~\cite{Pine-Nat-Phys},
true hyperuniformity has been reported numerically by Hexner and 
Levine~\cite{levine}
and experimentally by Weijs \emph{et al.}~\cite{bartolo-hyperuniform}
with hyperuniform exponent $\lambda \approx 0.45$.
Here we report that the value of $\lambda=0.45$ is only a crossover value found at intermediate range of $q$-vector
and we report an even stronger hyperuniformity  
with exponent $\lambda=1$~\cite{elsen1}.

The total structure factor is defined as:
\begin{equation}
S(q)=\left\langle \frac{1}{N}\sum_{i=1}^{N}\sum_{j=1}^{N}e^{i\mathbf{q}\cdot(\mathbf{r}_{i}-\mathbf{r}_{j})}\right\rangle -\left\langle N\right\rangle \delta_{\mathbf{q},0}. \label{eq:Sq}
\end{equation}
A signature of hyperuniform scaling is the behaviour of the total structure factor at low $q$:
\begin{equation}
S(q)\propto q^{\lambda}, \quad  q\rightarrow0.\label{eq:Sq-hyperuniform}
\end{equation}
since the limit $q\rightarrow0$ of the structure factor $S(q)$ is related to the number density fluctuation squared:
$S(q\rightarrow0)=\left\langle \Delta N^{2}\right\rangle / \left\langle N\right\rangle$.

\begin{figure} 
\begin{centering}
\includegraphics[width=1\columnwidth]{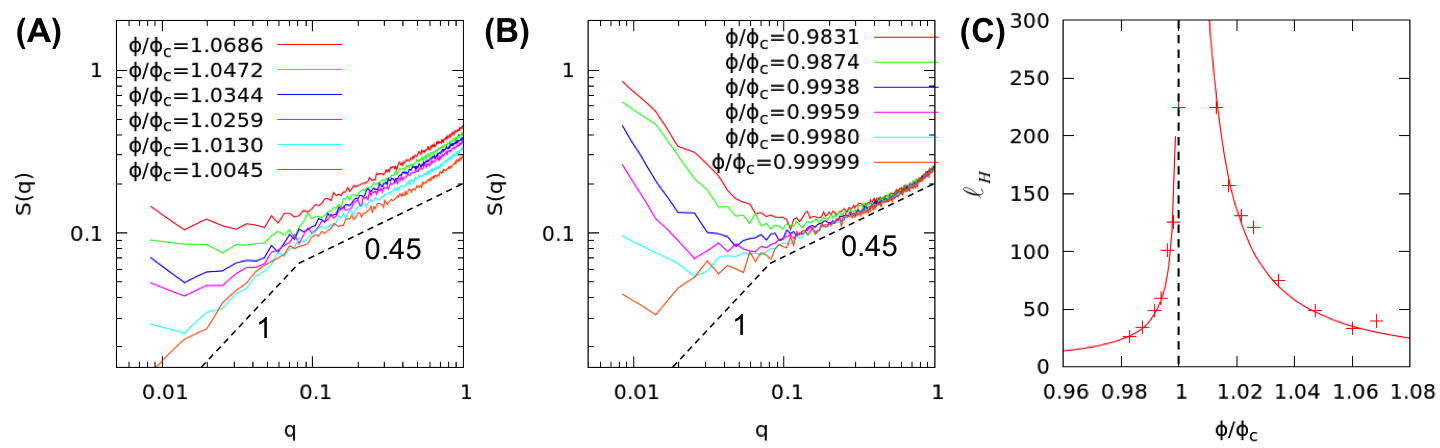}
\par\end{centering}
\caption{
Hyperuniform scaling close to criticality $\phi_c$. (The system size used 
in this case is $L=560$.)
\textbf{(A)} 
Total structure factor $S(q)$ for different densities as we 
approach the critical density from above $\phi\rightarrow\phi_{c}^{+}$.
Close to criticality, we detect a clear signature of hyperuniformity 
$S(q)\propto q^\lambda$ with hyperuniform exponent $\lambda=0.45$
crossing over to $\lambda=1$ at lower $q$.
\textbf{(B)}
Total structure factor as we approach the critical density 
from below $\phi\rightarrow\phi_{c}^{-}$.
Again we observe hyperuniform scaling with exponent $0.45$ crossing 
over to $1$. 
\textbf{(C)}
Hyperuniform length scale $\ell_H$ defined as the largest 
system size below which hyperuniform scaling holds for a given density $\phi$.
It appears to diverge asymmetrically from both sides of $\phi_c$,
suggesting that a true hyperuniform structure exists precisely 
at criticality.}
\label{fig:hyperuniform}
\end{figure}

We show in 
Figs.~\ref{fig:hyperuniform}(A,B) the total structure factor $S(q)$
as a function of $q$ for different densities 
above and below the critical density $\phi_{c}$. 
Approaching criticality from above $\phi\rightarrow\phi_c^+$ (Fig.~\ref{fig:hyperuniform}(A)),
we observe a clear signature of hyperuniform scaling: $S(q)\propto q^\lambda$ with hyperuniform exponent $\lambda=0.45$
crossing over to $\lambda=1$ at lower values of $q$.
Note that the intermediate value of $\lambda=0.45$ is the one that has been reported in Ref.~\cite{levine}.
Similarly approaching criticality from below 
$\phi\rightarrow\phi_c^-$ (Fig.~\ref{fig:hyperuniform}(B)),
we also observe a clear hyperuniform scaling with 
exponent $\lambda=0.45$ crossing over to $\lambda=1$.
Note that for $\phi<\phi_{c}$,
we perform several simulations from different initial random configurations and wait until each run goes into a different absorbing state.
The static structure factor is then ensemble-averaged over different runs.
For $\phi>\phi_c$, long time average at steady state is sufficient since the system is dynamically exploring its available phase space. 

Finally, following Ref.~\cite{bartolo-hyperuniform}, 
we may define a hyperuniform length scale $\ell_H(\phi)$ to be the largest system size below which hyperuniform scaling holds for a given density $\phi$.
$\ell_H$ can be obtained from the smallest value of $q$-vector at which the structure factor $S(q)$ starts to deviate from the hyperuniform scaling: $S(q)\sim q^\lambda$.
We find that the hyperuniform length scale tends to diverge as we approach criticality
both from below and above $\phi\rightarrow\phi_c^\pm$ 
(see Fig.~\ref{fig:hyperuniform}(C)). 
This implies that true hyperuniformity might survive in the large system size limit at $\phi_c$.
By fitting power law: $\ell_H\sim|\phi-\phi_c|^{-\nu_H}$ we find the critical exponent for $\ell_H$ to be $\nu_H\simeq0.76$ for $\phi\rightarrow\phi_c^-$ and $\nu_H\simeq1.23$ for $\phi\rightarrow\phi_c^+$.
Note that the critical exponent of the hyperuniform lengthscale $\nu_H$ 
is different to those of static and dynamic lengthscales $\nu_\perp$, 
and also appears to be different on both sides of the transition. 
One reason explaining this asymmetry lies in the fact that the final 
structure of absorbing states below $\phi_c$ crucially depends on the 
initial configurations, whereas above $\phi_c$ the phase space 
is explored dynamically in a nonequilibrium steady state. 
A recent numerical work has shown that 
the configurations explored by the dynamics below $\phi_c$
differ from an equilibrium sampling of the available phase 
space~\cite{frenkel-hyperuniform}. (Note that the work
of Ref.~\cite{frenkel-hyperuniform} would have been technically much 
easier--but conceptually equivalent--using 
our isotropic version of the model by Cort\'e {\it et al.}, since 
it would only involve canonical hard sphere simulations.)

\subsection{Pair distribution functions \label{sec:pair-dist}}

The pair distribution function between active-active or passive-passive particles is defined to be \cite{bhatia-thornton}:
\begin{equation}
g_{\alpha\alpha}(r)= \frac{1}{\left\langle n_{\alpha}
\right\rangle} \left\langle \sum_{i=1}^{N_{\alpha}}
\sum_{j \neq i}
\delta(\mathbf{r}-(\mathbf{r}_{i}-\mathbf{r}_{j}))\right\rangle, \label{eq:pair-dist-function}
\end{equation}
where $\alpha=a,p$ refers to active or passive particles respectively.
Physically, $\langle n_{\alpha} \rangle g_{\alpha\alpha}(r)$ tells us the average density of $\alpha$ particles at a distance $r$ from a given particle of the same species $\alpha$. 
Note that for isotropic and homogenous systems, $g_{\alpha\alpha}(r)$ only depends on the radial distance $r$. 
Furthermore we also have the asymptotic limit: $g_{\alpha\alpha}(r\rightarrow\infty)=1$, 
since the particles become uncorrelated at large distance. 
It can be easily shown that $g_{\alpha\alpha}(r)$ is related to the inverse Fourier transform of the structure factors $S_{\alpha\alpha}(q)$ defined in Eq.~(\ref{eq:structure-factor-2}):
\begin{equation}
h_{\alpha\alpha}(r)=g_{\alpha\alpha}(r)-1=\frac{1}{\left\langle n_{\alpha}\right\rangle }\frac{1}{(2\pi)^{2}}\int d\mathbf{q}\, e^{-i\mathbf{q}\cdot\mathbf{r}}\left\{ S_{\alpha\alpha}(q)-1\right\}, \label{eq:FT-pair-dist-function}
\end{equation}
where we have defined, for convenience, $h_{\alpha\alpha}(r)=g_{\alpha\alpha}(r)-1$.
Similarly, the mixed pair distribution function between active and passive particles is defined to be:
\begin{equation}
g_{pa}(r)=\frac{1}{\sqrt{\left\langle n_{p}\right\rangle \left\langle n_{a}\right\rangle }}\left\langle \sum_{i=1}^{N_{a}}\sum_{j=1}^{N_{p}}\delta(\mathbf{r}-(\mathbf{r}_{i}^{a}-\mathbf{r}_{j}^{p}))\right\rangle. \label{eq:mixed-pair-dist-function}
\end{equation}
Physically, $\langle {n}_{p}\rangle g_{pa}(r)$ gives us the average density of passive particles at a distance $r$ from a given active particle
(and \emph{vice versa} since $g_{pa}(r)=g_{ap}(r)$). 
Again, the mixed pair distribution function is related to the inverse Fourier transform of the mixed structure factor $S_{pa}(q)$ defined in Eq.~(\ref{eq:mixed-structure-factor}):
\begin{equation}
h_{pa}(r)=g_{pa}(r)-1=\frac{1}{\sqrt{\left\langle n_{p}\right\rangle \left\langle n_{a}\right\rangle }}\frac{1}{(2\pi)^{2}}\int d\mathbf{q}\, e^{-i\mathbf{q}\cdot\mathbf{r}}S_{pa}(q). \label{eq:FT-mixed-pair-dist-function}
\end{equation}

\begin{figure} 
\begin{centering}
\includegraphics[width=1\columnwidth]{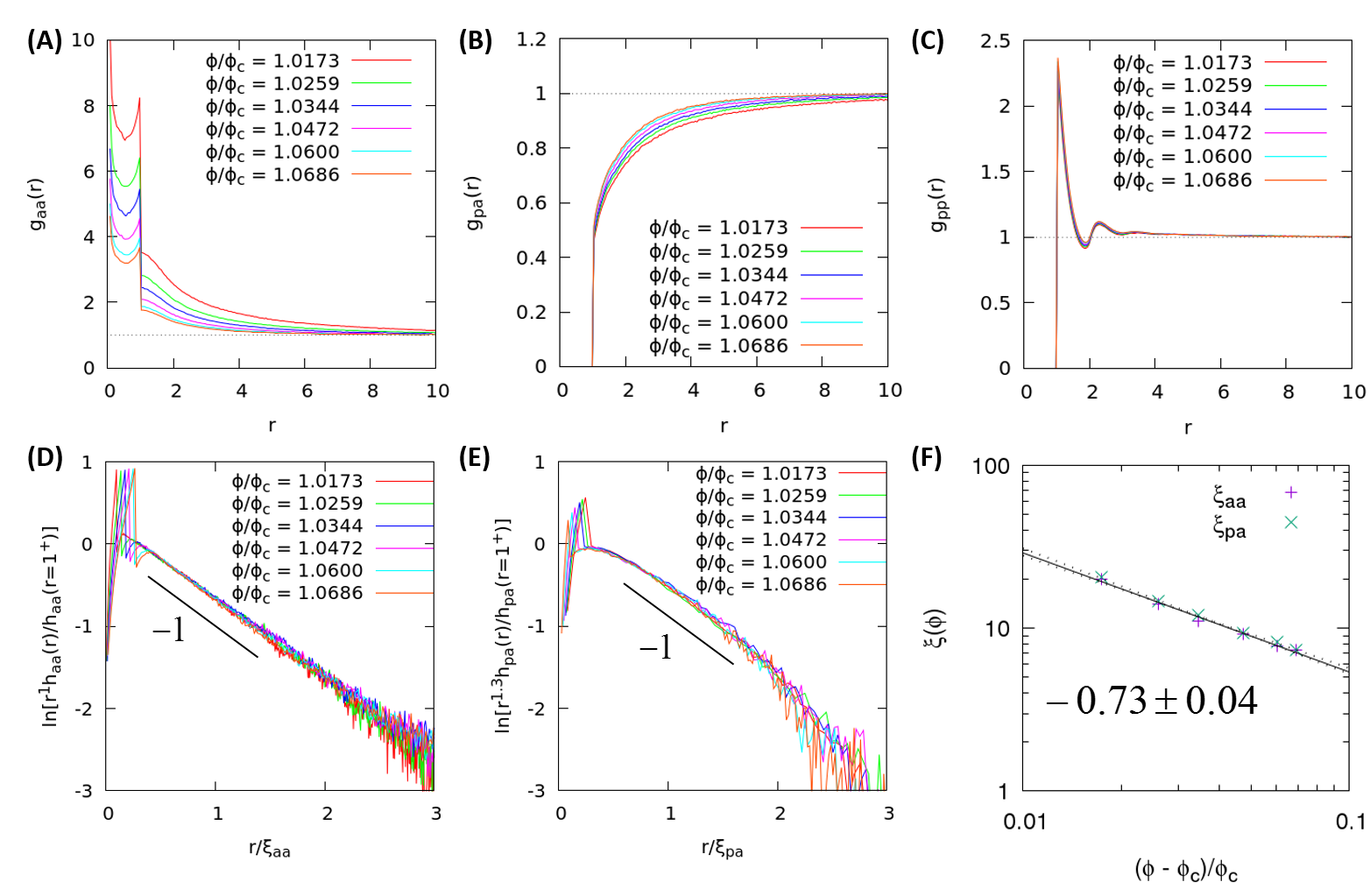}
\par\end{centering}
\caption{
\textbf{(A,B,C)}
Pair distribution functions for active-active particles $g_{aa}(r)$, 
passive-active particles $g_{pa}(r)$ and 
passive-passive particles $g_{pp}(r)$. 
\textbf{(D,E)}
The pair distribution functions can be rescaled by the appropriate 
lengthscales $\xi_{\alpha\beta}(\phi)$
(see text for the scaling form used to fit the functions). 
\textbf{(F)} The
extracted lengthscales for active-active particles (red plus)
and passive-active particles (green cross)
diverge as $\xi\sim(\phi-\phi)^{-\nu_\perp}$
with $\nu_\perp\simeq0.73$.}
\label{fig:pair-dist-functions}
\end{figure}

For active-active particles, the pair distribution function $g_{aa}(r)$ is plotted in Fig.~\ref{fig:pair-dist-functions}(A).
As can be seen from the figure, $g_{aa}(r\rightarrow\infty)=1$ as expected. 
However the value of $g_{aa}(r\simeq1)$ diverges as $\phi\rightarrow\phi_{c}^{+}$. 
This suggests clustering of active particles as we approach critical density. 
Furthermore we have verified that $g_{aa}(r\simeq1)$ scales as $1/\langle n_a \rangle$, 
so that the density of the active particles inside the cluster ($=\langle n_{a} \rangle g_{aa}(r=1)$)
is independent of the density of the whole system. 
This is as we expect because the active particles inside the cluster behave like an ideal gas, and consequently, 
the density of the cluster only depends on the size of the random kick $\delta$. 
Similarly from the definition of passive-active pair distribution function, Eq.~(\ref{eq:mixed-pair-dist-function}),
we expect $g_{pa}(r\simeq1)$ to scale as $1/\sqrt{\langle n_{a} \rangle \langle n_{p}\rangle }$.
The pair distribution function for the active-active, passive-active and passive-passive are plotted in 
Figs.~\ref{fig:pair-dist-functions}(A,B,C) respectively. 
The pair distribution function for the passive-passive case (Fig.~\ref{fig:pair-dist-functions}(C)) resembles somewhat that of hard spheres system. 
This is because all passive particles do 
not overlap with each other, thus behaving qualitatively similarly
to a hard sphere fluid, with quantitative 
differences~\cite{frenkel-hyperuniform}.

From the snapshots in Fig.~\ref{fig:snapshot-static}, we see a growing static correlation length as we approach criticality. 
This is reflected both in the structure factors (as we have seen from previous subsection) and in the pair distribution function. 
In the case of the pair distribution function, we expect $g_{\alpha\beta}(r)$ to decay as 
$g_{\alpha\beta}(r)\sim e^{-r/\xi_{\alpha\beta}}$ for large $r$
where $\xi_{\alpha\beta}$ is the static correlation length. 
For instance, $\xi_{aa}$ gives us the typical size of the active clusters and $\xi_{pa}$ gives us the typical distance between the clusters. 
Subsequently close to criticality, we may assume the following asymptotic form for the pair distribution function:
\begin{equation}
\frac{h_{\alpha\beta}(r)}{h_{\alpha\beta}(r=1^{+},\phi)}\simeq\frac{1}{r^{\theta'_{\alpha\beta}}}e^{-\frac{r}{\xi_{\alpha\beta}(\phi)}},\quad {\rm for} \,\,\, r\gg\xi,
\label{eq:rescaled-pair-dist-function}
\end{equation}
where we know from above, $h_{\alpha\beta}(r=1^{+})$ scales as $1/\langle n_{a}\rangle $ for the active-active case
and $1/\sqrt{\langle n_{a}\rangle \langle n_{p}\rangle }$ for the passive-active case. 
The exponent $\theta'$ in the equation above can be derived from the structure factors as follows. 
First we recognize that the asymptotic limit $r\gg\xi$ is equivalent to the limit $q\ll\frac{2\pi}{\xi}$ in the Fourier space. 
For example, the active-active pair distribution function $h_{aa}(r)$ can be obtained from the Fourier transform of the structure factor as follows:
\begin{eqnarray}
S_{aa}(q) & = & 1+\left\langle n_{a}\right\rangle \int d^{2}r\, e^{i\mathbf{q}\cdot\mathbf{r}}h_{aa}(r)\nonumber \\
 & \simeq & 1+A\int d^{2}r\, e^{i\mathbf{q}\cdot\mathbf{r}}\frac{1}{r^{\theta'_{aa}}}e^{-\frac{r}{\xi_{aa}(\phi)}}\quad {\rm for } \,\,\, 
q\ll\frac{2\pi}{\xi_{aa}}, 
\label{eq:Saa-1}
\end{eqnarray}
where $A=\left\langle n_{a}\right\rangle h_{aa}(r=1^{+})$ is some constant which does not depend on $\phi$ (since $h_{aa}(r=1^{+})$ scales as $1/\langle n_{a}\rangle $). 
Finally by change of variable: $u=r/\xi_{aa}$, we have for small $q\ll\frac{2\pi}{\xi_{aa}}$:
\begin{eqnarray}
S_{aa}(q) & \simeq & 1+\xi_{aa}(\phi)^{2-\theta'_{aa}}\underbrace{A\int d^{2}u\, e^{i\mathbf{q}\xi_{aa}(\phi)\cdot\mathbf{u}}\frac{1}{u^{\theta'_{aa}}}e^{-u}}_{J(q\xi_{aa}(\phi))}, 
\quad{ \rm for } \, \,\,  q\ll\frac{2\pi}{\xi_{aa}}. 
\label{eq:Saa-2}
\end{eqnarray}
where $J(x)$ is a universal function which does not depend on $\phi$
(compare to Eq.~(\ref{eq:rescaled-structure-factor})). Therefore
in the limit $q\rightarrow0$, we have
\begin{equation}
S_{aa}(q\rightarrow0)\simeq1+\xi_{aa}(\phi)^{2-\theta'_{aa}}J(0),
\label{eq:theta-prime}
\end{equation}
and comparing this equation to Eq.~(\ref{eq:theta}), we deduce that $\theta'_{aa}=\theta_{aa}$ 
(similarly for the passive-active case $\theta'_{pa}$). 
\revise{Imposing the values} of $\theta_{aa}=1$ and $\theta_{pa}=1.3$ obtained from the structure factors, 
we may collapse the pair distribution functions into the 
scaling form in Eq.~(\ref{eq:rescaled-pair-dist-function}),
shown below, suggesting that our measurements are 
\revise{at least internally consistent. These measurements 
cannot provide a very accurate estimate of the exponent $\theta_aa$
to be used in the distinction between DP and CDP universality classes.}

In Fig.~\ref{fig:pair-dist-functions}(D), we plot $\ln\left[r\frac{h_{aa}(r)}{h_{aa}(r=1^+,\phi)}\right]$ as a function of $\frac{r}{\xi_{aa}(\phi)}$. 
By choosing the appropriate values for $\xi_{aa}(\phi$), 
we can collapse all pair distribution functions for different $\phi$'s into a single straight line with slope $-1$. 
From this universal scaling, we extract a lengthscale $\xi_{aa}(\phi)$ which has power law divergence close to $\phi_{c}$
(see Fig.~\ref{fig:pair-dist-functions}(F)). 
Similarly for the mixed passive-active case, we plot $\ln\left[r^{1.3}\frac{h_{pa}(r)}{h_{pa}(r=1^+,\phi)}\right]$ as a function of $\frac{r}{\xi_{pa}(\phi)}$ 
(see Fig.~\ref{fig:pair-dist-functions}(E)).
From the data collapse, we extract the lengthscale $\xi_{pa}(\phi)$ which is plotted in Fig.~\ref{fig:pair-dist-functions}(F). 
Assuming both lengthscales are equivalent, we find the critical exponent for the lengthscale to be: $\nu_\perp\simeq0.73$ (see Fig.~\ref{fig:pair-dist-functions}(F)).
Note that this value of the critical exponent is indeed consistent to the one obtained from the structure factors in Sec.~\ref{sec:structure-factors}.
Finally it should be mentioned that, for the passive-passive case,
the values of the pair distribution function for different $\phi$'s are too close to each other and no good statistics could be obtained to extract the lengthscale.

\section{Dynamic properties and heterogeneities}

\label{sec:dynamic}

\subsection{Single particle dynamic heterogeneities 
\label{sec:single_particle_dynamics}}

The dynamics of the system at the particle scale can be characterised by two timescales.
1) The timescale for an active particle to become passive ($\tau_{p}$) and 
2) the timescale for a passive particle to become active ($\tau_{a}$). 
The passivation timescale $\tau_{p}$ is typically much shorter and does not depend on the packing fraction.
In fact, the passivation timescale only depends on the size of the random displacement $\delta$. 
On the other hand the activation timescale $\tau_{a}$ becomes longer and diverges as we approach the critical packing density from above ($\phi\rightarrow\phi_{c}^{+}$).

\begin{figure} 
\begin{centering}
\includegraphics[scale=0.7]{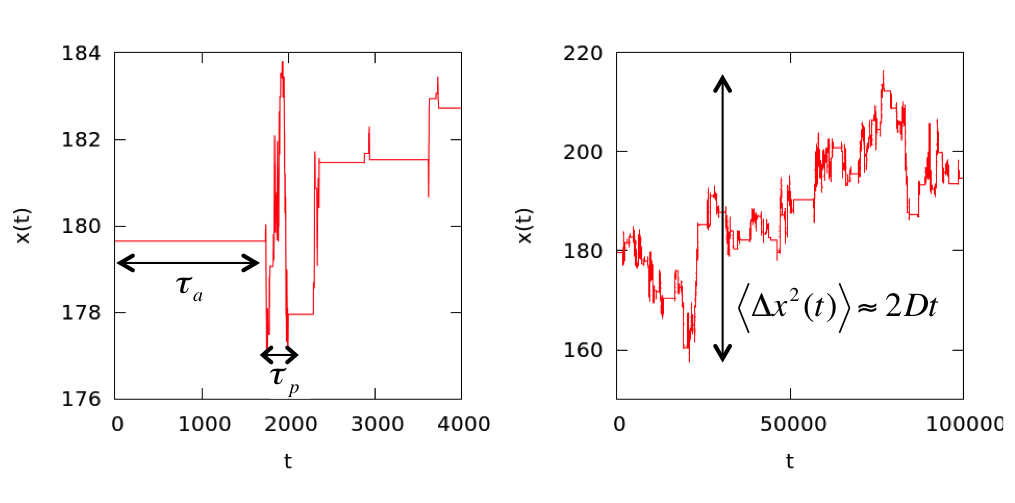}
\par\end{centering}
\protect\caption{
Typical trajectory of a single particle in the system (projected along the $x$-axis). 
The particle remains passive 
during some time interval characterised by $\tau_{a}$ and 
becomes mobile during another 
time interval characterised by $\tau_{p}\ll\tau_{a}$. 
If we look at the trajectory of the particle over larger times 
compared to $\tau_a$ and $\tau_p$, as in the right panel,
the trajectory becomes typical of an ordinary random 
walker.}
\label{fig:activation-passivation}
\end{figure}

Fig.~\ref{fig:activation-passivation} shows the typical trajectory of a single particle (in the $x$-direction). 
From the trajectory, we can see that the particle remains stationary (passive) for some time before moving (active) and so on. 
The activation timescale $\tau_{a}$ is thus the typical timescale for a single particle to remain stationary/passive. 
Similarly, the passivation timescale $\tau_{p}$ is the typical timescale for the particle to remain mobile/active (see left panel in Fig.~\ref{fig:activation-passivation}). 
(This is analogous to the trajectory of the particles in glassy systems where 
the particle is caged by their neighbours and there exists an activation timescale for the particle to escape the cage \cite{berthier-fickian}.) 
Over large distances and times as compared to $\tau_a$, 
the trajectory of the particle eventually becomes indistinguishable from that of a random walker (see right panel of Fig.~\ref{fig:activation-passivation}). 
Thus at large distances and times (as compared to the activation timescale), 
all particles are diffusive with diffusion constant $D$:
\begin{equation}
4D=\lim_{t\rightarrow\infty}\frac{1}{t}\left\langle \frac{1}{N}\sum_{i=1}^{N}\left|\Delta\mathbf{r}_{i}(t)\right|^{2}\right\rangle, \label{eq:D}
\end{equation}
where $\Delta\mathbf{r}_{i}(t)=\mathbf{r}_{i}(t'+t)-\mathbf{r}_{i}(t')$
with $t'$ larger than the steady state time $t'>t_{ss}$. 
The angle bracket indicates time average over many initial times $t'$ as before. 
Note that the diffusion constant is a single particle statistics. 
Naturally, we expect the diffusion constant to be zero in the passive phase since 
the number of overlaps (hence mobile or active particles) goes to zero at steady state. 
On the other hand, the diffusion constant is finite in the active phase, \emph{i.e.}
\begin{equation}
D(\phi) = 0, \,\,\, \phi<\phi_{c}, \quad \quad \quad \quad 
D(\phi) > 0, \,\,\, \phi>\phi_{c}.
\label{eq:D_phi}
\end{equation}
Moreover, it can be shown that $D$ is proportional to the order parameter/fraction of active particles in the system: $D\propto\langle f_{a}\rangle$.
To see this, we observe that
\begin{equation}
\Delta\mathbf{r}_{i}(t)=\int_{t'}^{t'+t}dt''\, f_{i}(t'')\vec{\delta}_{i}(t''),
\label{eq:D_r}
\end{equation}
where $\vec{\mathbf{\delta}}_{i}$ is the random displacement that we give to particle $i$ (but only if particle $i$ is active, \emph{i.e. $f_{i}=1$}). 
From Eq.~(\ref{eq:D_r}), it follows that $\left|\Delta\mathbf{r}_{i}(t)\right|^{2}=\left\langle f_{a}\right\rangle \langle \vec{\delta}_{i}^{2}\rangle t$ and thus:
\begin{equation}
D=\left\langle f_{a}\right\rangle D_{a}, \label{eq:D-1}
\end{equation}
where $D_{a}=\langle \vec{\delta}_{i}^{2}\rangle /4$ is the diffusion 
constant that an (hypothetical) \emph{always}-active particle would have. 
Finally since $D\propto\langle f_{a}\rangle $, 
the diffusion constant vanishes continuously at critical point with the same critical exponent as that of the order parameter 
($D\sim(\phi-\phi_{c})^{\beta}$ with $\beta\simeq0.59$).
Effectively, the diffusion constant does not give us new information but experimental measurement of $D(\phi)$
can be useful to infer $\left\langle f_{a}\right\rangle (\phi)$.

\begin{figure} 
\begin{centering}
\includegraphics[scale=0.7]{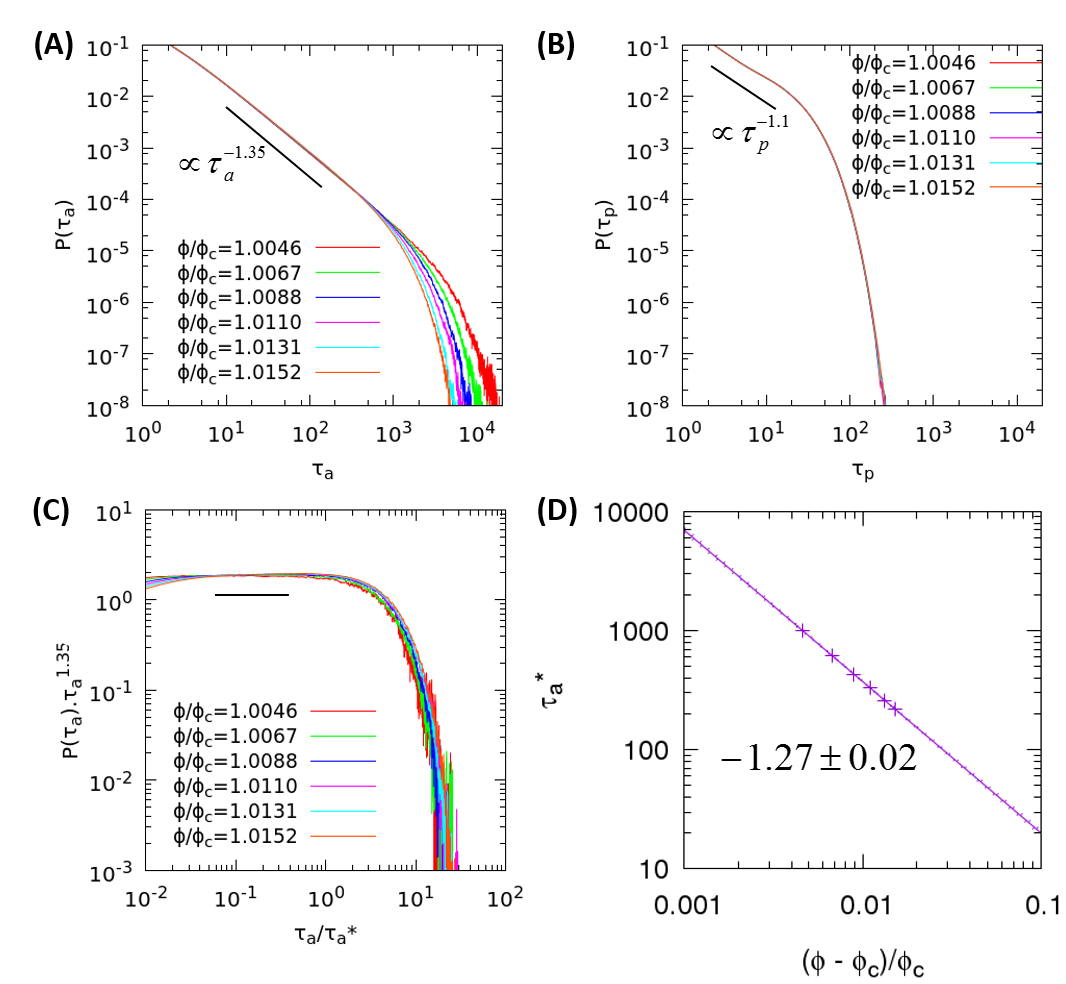}
\par\end{centering}
\caption{
\textbf{(A)}
The distribution of the activation timescale $P(\tau_{a})$ exhibits 
a power law behaviour at short times and an exponential cut-off at large times: 
$P(\tau_{a})\sim\tau_{a}^{-z}\exp\left(-
\tau_{a} / \tau_{a}^{*}(\phi) \right)$ with $z=1.35$
and an increasing timescale $\tau_{a}^{*}(\phi)$ as 
$\phi\rightarrow\phi_{c}^{+}$. 
\textbf{(B)}
The distribution of the passivation timescale $P(\tau_{p})$ 
also exhibits a power law behaviour with an exponential cut-off. 
Contrary to $P(\tau_{a})$, $P(\tau_{p})$ does not depend on the density.
\textbf{(C)} 
To find $\tau_{a}^{*}(\phi)$, we collapse $\tau_{a}^{z}P(\tau_{a})$ 
against $\tau_{a}/\tau_a^*(\phi)$.  
\textbf{(D)}
The characteristic timescale $\tau_{a}^{*}(\phi)$ diverges 
as $\tau_{a}^{*}\sim(\phi-\phi_{c})^{-1.25}$.}
\label{fig:tau-dist}
\end{figure}

Figs.~\ref{fig:tau-dist}(A,B) show the distribution of the activation and passivation timescales. 
As can be seen from the logarithmic scale, both distributions exhibit a power law behaviour at short times with an exponential cut-off at large times, \emph{i.e.}
\begin{equation}
P(\tau)\sim\tau^{-z}e^{-\tau/\tau^{*}(\phi)}. \label{eq:tau-dist}
\end{equation}
In the case of the distribution of activation timescales $P(\tau_{a})$,
the exponential tail becomes longer as we approach the critical point
(see Fig.~\ref{fig:tau-dist}(A)). 
On the other hand, the probability distribution of the passivation timescales $P(\tau_{p})$ does not change with the packing fraction $\phi$ since 
$\tau_{p}$ only depends on the size of the random kicks $\delta$ that we give to the active particles (see Fig.~\ref{fig:tau-dist}(B)). 
{\color{black}
Furthermore at steady state, we have $\langle f_a \rangle \langle \tau_a \rangle = (1-\langle f_a \rangle) \langle \tau_p \rangle$ due to equal flux of passive to active and passive to active conversion, where $\langle \tau_{a,p} \rangle$ is the average of $\tau_{a,p}$ over the probability distribution $P(\tau_{a,p})$.}

We can characterise the exponential cutoff by some relaxation timescale $\tau_a^{*}$ according to Eq.~(\ref{eq:tau-dist}). 
In the case of the $P(\tau_{a})$, this time scale $\tau_{a}^{*}$ grows as we approach the critical density $\phi\rightarrow\phi_{c}^{+}$. 
To see this, we plot $\tau_{a}^{z}P(\tau_{a})$ as a function of $\tau_{a}/\tau_a^*(\phi)$ 
such that all plots for different $\phi$'s collapse into a single curve at large $\tau_{a}$ (see Fig.~\ref{fig:tau-dist}(C)). 
From data collapse, we obtain $\tau_{a}^{*}$ as a function of the packing fraction $\phi$ which indeed diverges at criticality with a power law behaviour:
$\tau_{a}^{*}\sim(\phi-\phi_{c})^{-\nu_\parallel}$ as $\phi\rightarrow\phi_{c}^{+}$ with critical exponent $\nu_\parallel\simeq1.25$
(see Fig.~\ref{fig:tau-dist}(D)). 
Below we shall look at how the characteristic activation timescale $\tau_{a}^{*}$ may be probed experimentally.

\begin{figure} 
\begin{centering}
\includegraphics[width=0.7\columnwidth]{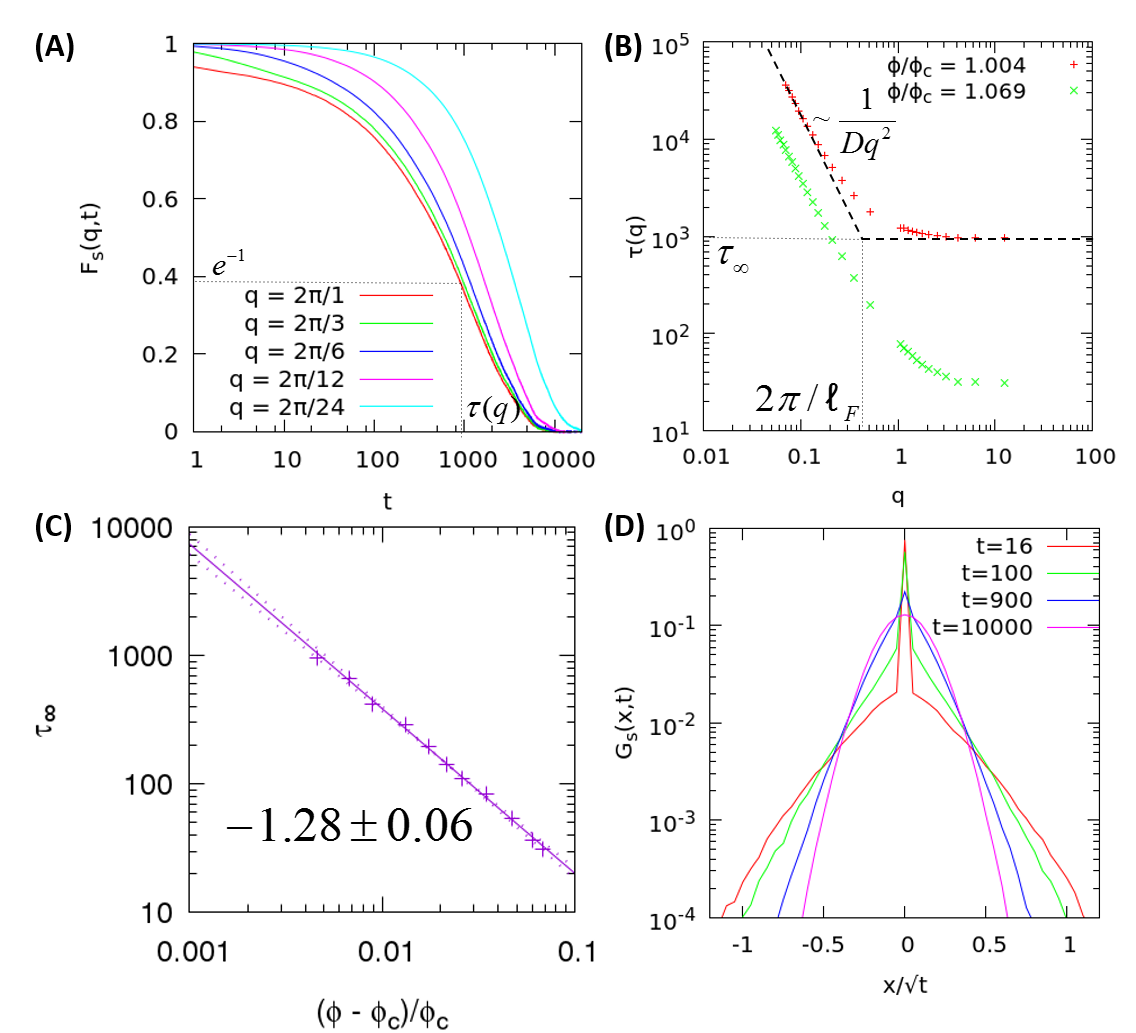}
\par\end{centering}
\caption{
\textbf{(A)}
Intermediate scattering function 
as a function of time $t$ for different wavectors 
$q$ for density $\phi/\phi_c=1.0044$.
\textbf{(B)} The
wave-vector dependence of relaxation
time $\tau(q)$ for two different densities
crosses over from diffusive when $q \rightarrow 0$,  
and becomes strongly non-Fickian and saturates to a constant value 
$\tau_\infty$ at large 
$q$. The crossover from Fickian to non-Fickian dynamics
defines a Fickian lengthscale $\ell_F(\phi)$. 
\textbf{(C)} 
The timescale $\tau_{\infty}$ diverges 
at $\phi_c$ with critical exponent
$\nu_\parallel \simeq 1.28$.
\textbf{(D)} 
The distribution of particle displacements 
$G_{s}(x,t)$ displays a sharp peak at $x=0$
for small times and tails that are broader than a Gaussian.
The distribution converges to a Gaussian distribution at large times.}
\label{fig:tau-q}
\end{figure}

Let us define $c(q,t)$ as:
\begin{equation}
c(q,t)=\frac{1}{N}\sum_{i=1}^{N}e^{i\mathbf{q}\cdot\Delta\mathbf{r}_{i}(t)},
\label{eq:c_q_t}
\end{equation}
where $\Delta\mathbf{r}_{i}(t)=\mathbf{r}_{i}(t'+t)-\mathbf{r}_{i}(t')$
is the particle's displacement after time $t$. 
Physically, the quantity above tells us about the fraction of particles which have moved a distance 
$\lambda=\frac{2\pi}{q}$ during the time interval $[t',t'+t]$.
The intermediate scattering function is defined to be the average
of $c(q,t)$ over many initial times $t'$:
\begin{equation}
F_{s}(q,t)=\left\langle c(q,t)\right\rangle. \label{eq:Fs_q_t}
\end{equation}
Experimentally, $F_{s}(q,t)$ is obtained from the intensity of monochromatic light scattered from the sample, or directly using microscopy techniques 
depending on the typical particle size considered. 
Fig.~\ref{fig:tau-q}(A) shows the typical intermediate scattering function as a function of time $t$
for different wavevectors $q$. 
By equating $F_{s}(q,\tau(q))=e^{-1}$, we obtain a relaxation timescale $\tau(q)$ which depends on the wavevector $q$ (see Fig.~\ref{fig:tau-q}(B)). 
From this figure, we see that $\tau(q)$ scales as $q^{-2}$ 
for small $q$ which is expected
since in the limit of $q\rightarrow0$ (long wavelengths and timescales limit), 
all particles behave like a random walker (see the trajectory in Fig.~\ref{fig:activation-passivation}(B) right). 
On the other hand for large $q\rightarrow\infty$, 
the relaxation time saturates to a finite value $\tau_{\infty}=\tau(q\rightarrow\infty)$. 
Thus from the intermediate scattering function, we may extract a characteristic
timescale $\tau_{\infty}(\phi)$ which diverges as we approach criticality from above
$\phi\rightarrow\phi_{c}^{+}$ with scaling law: $\tau_{\infty}\sim(\phi-\phi_{c})^{-\nu_{\parallel}}$
and a critical exponent $\nu_{\parallel}\simeq1.28$ (see Fig.~\ref{fig:tau-q}(B)). 

Physically, $\tau_{\infty}$ gives us the timescale for all particles
to move from their original positions $\{\mathbf{r}_{i}(t')\}$. 
Thus, $\tau_{\infty}$ is also the waiting time for all particles to become
active at some point during the time interval $[t',t'+t]$. 
However, we also know that the waiting time $\tau_{\infty}$ is constrained
by the probability distribution of the activation timescales $P(\tau_{a})$.
In particular, $\tau_{\infty}$ is limited by the long tail distribution
of $P(\tau_{a})$. From Fig.~\ref{fig:tau-dist}(C), $P(\tau_{a})$
has an exponential tail as follows:
\begin{equation}
P(\tau_{a})\sim \exp \left( 
- \frac{\tau_{a}}{\tau{}_{a}^{*}(\phi)}  \right),
\label{eq:tau_act_dist}
\end{equation}
thus we deduce that $\tau_{\infty}\simeq\tau{}_{a}^{*}$. Indeed,
both $\tau_{\infty}(\phi)$ and $\tau_{a}^{*}(\phi)$ diverge as $\phi\rightarrow\phi_{c}^{+}$
with the same critical exponent (\emph{cf.} Fig.~\ref{fig:tau-dist}(D)
to Fig.~\ref{fig:tau-q}(C)). 

Finally, we may define a Fickian/diffusion lengthscale to be: $\ell_F=\sqrt{4D\tau_\infty}$.
This lengthscale is defined to be the crossover lengthscale from a Fickian [$\tau\sim1/(Dq^2)$] 
to a non-Fickian dynamics [$\tau\sim\tau_\infty(\phi)$] (see Fig.~\ref{fig:tau-q}(B)).
From above, we know that the diffusion constant and the relaxation time scale as: 
$D\sim(\phi-\phi_c)^{0.59}$ and $\tau_\infty\sim(\phi-\phi_c)^{-1.28}$ respectively. 
Thus the Fickian lengthscale should diverge as $\ell_F\sim(\phi-\phi_c)^{-0.35}$ 
as $\phi\rightarrow\phi_{c}^{+}$. 
Notice that it has a different critical exponent from the correlation length 
($\xi\sim(\phi-\phi_c)^{-0.73}$)
which we obtained in Sec.~\ref{sec:structure-factors} and Sec.~\ref{sec:pair-dist}.
This is as we should expect because the Fickian lengthscale is the average distance that a single particle diffuse (irrespective of other particles) during 
a correlation time.
On the other hand the correlation length $\xi$ contains information about the collective behaviour (such as clustering). The independance between 
the Fickian lengthscale and other correlation lenghtscales is also
found in supercooled liquids~\cite{berthier-fickian,berthier04}.

To illustrate further the onset of Fickian dynamics, we can look at
the distribution of the particles' displacements during the time interval
$[t',t'+t]$. More precisely, the van Hove function is defined to
be:
\begin{equation}
G_{s}(x,t)=\left\langle \frac{1}{N}\sum_{i=1}^{N}\delta(x-\Delta x_{i}(t))\right\rangle, \label{eq:Gs_x_t}
\end{equation}
where $\Delta x_{i}(t)=x_{i}(t'+t)-x_{i}(t')$ is the particle's displacement
in the $x$-direction. The angle bracket $\left\langle .\right\rangle $
indicates time averaging over many initial times $t'$. Basically,
this gives us the probability of finding a particle with displacement
$x$ after time $t$ (note that since the system is isotropic, the
probability distribution in the $x$-coordinate is the same as that
in the $y$-coordinate). The van Hove function is plotted against
$x/\sqrt{t}$ in Fig.~\ref{fig:tau-q} for different values of
$t$. (Notice that $x$ is rescaled by $\sqrt{t}$ to fit curves at
large $t$ which may become diffusive \emph{i.e.} $x\sim\sqrt{t}$.)
At short times $t\sim\tau$, we see a sharp peak at $x=0$, indicating
that a large fraction of particles have not moved (\emph{i.e. }remained
passive during $[t',t'+t]$). As $t$ increases, the peak gradually
decreases as more particles become active and diplaced from their
original positions $\{x_{i}(t')\}$. Finally at long times $t\gg\tau$,
all particles become diffusive and the van Hove function becomes Gaussian
with a width $\propto\sqrt{t}$.
The coexistence at short times of a large number of immobile particles with a
small number of mobile ones is natural in the context 
of the transition to an absorbing state, because 
the concentration of active particles vanishes continuously 
as $\phi \to \phi_c^+$. This is also an important dynamic 
signature in systems near a glass transition~\cite{berthier-book}, 
where mobility becomes sparse as the glass transition is 
approached. 

\subsection{Collective dynamic heterogeneities \label{sec:collective-dynamics}}

\begin{figure} 
\begin{centering}
\includegraphics[width=1\columnwidth]{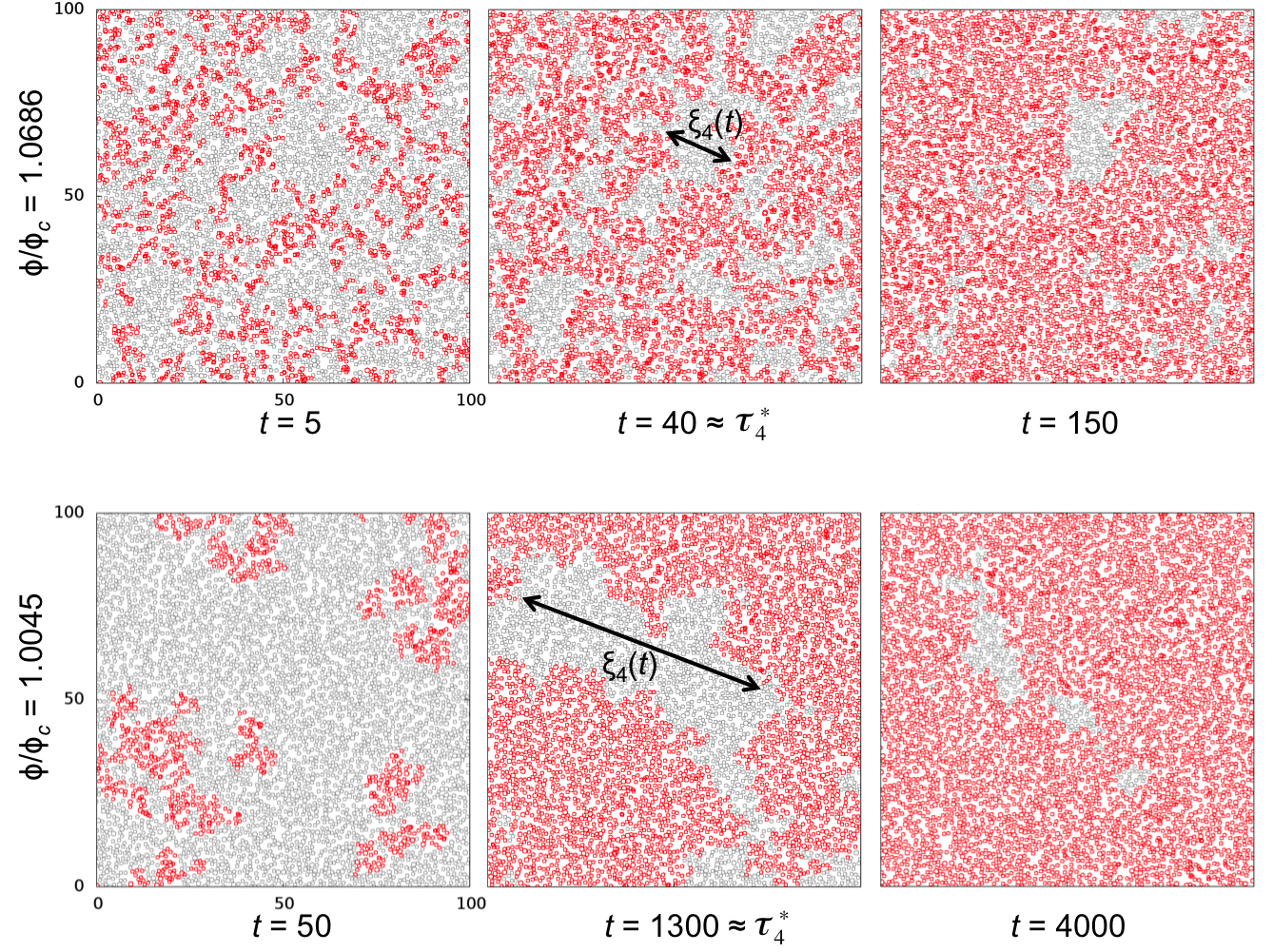}
\end{centering}
\caption{Time evolution of dynamic activity. Particles that 
have been mobile between the time interval $t$ are shown as red, 
always passive particles as grey.
The correlation length between mobile and immobile particles 
defines a dynamic lengthscale $\xi_{4}(t)$.
The dynamic lengthscale $\xi_{4}(t)$ increases with 
the time interval $t$ at fixed density, and also when approaching 
the critical point.
Note that the snapshots only show a small portion of a
larger system with $L=280$.}
\label{fig:snapshot-dynamic}
\end{figure}

In Sec.~\ref{sec:structure-factors}, 
we discussed how we may obtain a \emph{static} correlation length in our sytem  by looking at the distribution of the active and passive particles. 
By static, we mean that the activity of the particles is determined from one period of oscillation. 
In this section, we shall demonstrate how we can also obtain a dynamic lengthscale in our system. 
This means we are looking at the state/activity of the particles over time intervals larger than one period. 
To this end, we run the simulation for some time interval $[t',t'+t]$ where the initial time $t'$ is larger than the steady state time. 
During this time interval $[t',t'+t]$ we identify particles which have been active at least once 
(in other words, particles which have moved from their original positions at $\{\mathbf{r}_{i}(t')\}$).
We shall call these `A' particles (red colour in Fig.~\ref{fig:snapshot-dynamic}).
On the other hands, particles which remain passive or stay at the same positions during the whole time interval $[t',t'+t]$ are labelled P particles (grey colour in the figure). Let us insist again that these labels are defined over a finite
time interval and thus differ from the `$a$' and `$p$' labels used in 
Sec.~\ref{sec:static} above.

As we can see from Fig.~\ref{fig:snapshot-dynamic}, we observe domains of 
A particles which grow with the time interval $t$, reminiscent of 
spinodal decomposition. 
Eventually, the system will be filled with only A particles as 
$t\rightarrow\infty$. 
Thus we may define a dynamic lengthscale $\xi_{4}(t)$ to be the 
correlation length between A and P particles at time $t$. 
In Fig.~\ref{fig:snapshot-dynamic}, $\xi_{4}(t)$ represents the 
average distance between two clusters of A particles. 
Thus we expect the strength of dynamic correlations to 
increase initially before reaching a peak at $t=\tau_{4}*$~\cite{berthier-book}. 
Following recipes devised in studies 
of supercooled liquids, the dynamic correlation length is defined to be: 
$\xi_{4}^{*}=\xi_{4}(\tau_{4}^{*})$.
Comparing the two systems at two different densities in 
Fig.~\ref{fig:snapshot-dynamic}, both $\tau_{4}^{*}$ and $\xi_{4}^{*}$ 
grow larger as we approach the critical density 
$\phi\rightarrow\phi_{c}^{+}\simeq0.375$. 

We now discuss 
how this dynamic lengthscale $\xi_4$ and timescale 
$\tau_4^{*}$ may be quantified.
Recall the definition of the static structure factor $S(q)$ in Eq.~(\ref{eq:structure-factor-2}).
Similarly, we may also define the dynamic structure factor to be \cite{berthier-book}:
\begin{equation}
S_{4}(k,t)=\left\langle \frac{1}{N}\sum_{i=1}^{N}\sum_{j=1}^{N}e^{i\mathbf{k}\cdot(\mathbf{r}_{i}-\mathbf{r}_{j})}c_{i}(q,t)\left[c_{j}(q,t)-1\right]\right\rangle -\left\langle c_{i}(c_{i}-1)\right\rangle \delta_{\mathbf{k},0},
\label{eq:S4-mixed}
\end{equation}
where $c_{i}(q,t)$ is already defined in Eq.~(\ref{eq:c_q_t}). 
For the rest of the discussion below, we shall set the wavector $q$ to be fixed at: $q=2\pi$. 
(This value of $q$ corresponds to the saturation value of $\tau(q)$ in Fig.~\ref{fig:tau-q}).
Physically, $\langle c_{i}(q,t)\rangle =1$ if particle $i$ remains inactive during time interval $[t',t'+t]$ and $\langle c_{i}(q,t)\rangle =0$
if particle $i$ has been active at least once during time interval $[t',t'+t]$. 
Thus the dynamic structure factor defined above tells us about the correlation between A ($\left\langle c_{i}(q,t)\right\rangle =0$) and P particles ($\left\langle c_{i}(q,t)\right\rangle =1$)
in Fig.~\ref{fig:snapshot-dynamic} at time $t$ (this is analogous to $S_{pa}(q)$ in the static case). 

For completeness, we may also define the dynamic structure 
between pairs of A or P particles, which are the 
dynamical analogs to the partial structure
factors $S_{aa}(q)$ and $S_{pp}(q)$ in the static case.
However since we are only interested in the maximum fluctuations 
(which occurs at $t=\tau_{4}^{*}$), 
the distribution of A and P particles are almost symmetric 
(see middle panels in Fig.~\ref{fig:snapshot-dynamic}).
In fact we found that at $t=\tau_{4}^{*}$ and for small $k\ll2\pi/\xi_{4}$, 
all three partial dynamic structure factors are indeed similar. 
Thus for the rest of the discussion below, we shall only consider 
the mixed dynamic structure factor $S_{4}(k,t)$ as defined in 
Eq.~(\ref{eq:S4-mixed}).

\begin{figure} 
\begin{centering}
\includegraphics[width=1\columnwidth]{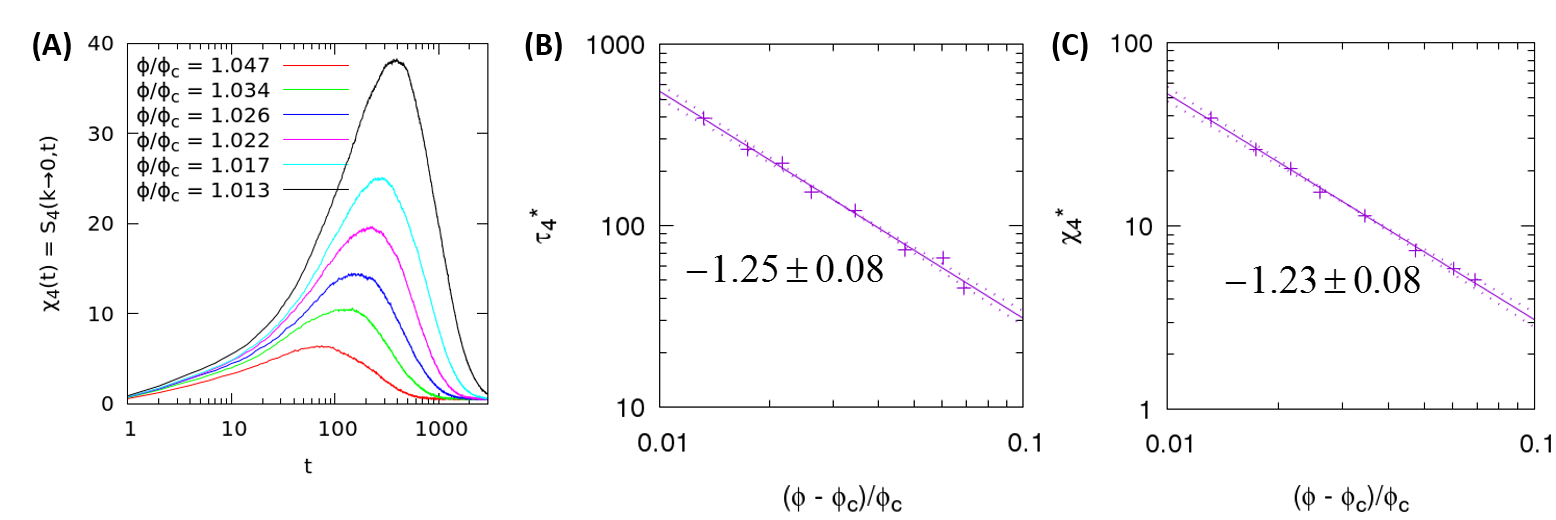}
\end{centering}
\caption{
\textbf{(A)} 
Time dependence of the 
dynamic susceptibility $\chi_{4}(t)$ quantifying 
correlation between particles which have moved (A particles) and
particles which are still at the same position (P particles) after time $t$. 
It shows a peak at $t=\tau_{4}^{*}(\phi)$, whose amplitude grows as 
$\phi \to \phi_c^+$.
\textbf{(B)}
The timescale $\tau_{4}^{*}$ diverges as
$\tau_{4}^{*}\sim(\phi-\phi_{c})^{-\nu_{\parallel}}$ where
$\nu_{\parallel} \simeq 1.25$. 
\textbf{(C)} The maximum correlation
$\chi_{4}^{*}$ also diverges at criticality,
$\chi_{4}^{*}\sim(\phi-\phi_{c})^{-\gamma'}$
where $\gamma' \simeq 1.23$. \label{fig:chi4}}
\end{figure}

We also define the corresponding dynamic susceptibility to be:
\begin{equation}
\chi_{4}(t)=S_{4}(k\rightarrow0,t)=\frac{\sqrt{\Delta N_{\rm A}(t)
\Delta N_{\rm P}(t)}}{N_{\rm A}(t)N_{\rm P}(t)}, \label{eq:chi4}
\end{equation}
which is the mixed number density fluctuation 
for A and P particles, analogous to the static 
case in Eq.~(\ref{eq:Spa}) from Sec.~\ref{sec:structure-factors}.
In other words, $\chi_{4}(t)$ measures the correlation between particles which have moved and particles which are still at the same positions after time $t$. 
The dynamic susceptibility is plotted in Fig.~\ref{fig:chi4}(A) for different densities $\phi$. 
As we can see from Fig.~\ref{fig:chi4}(A), $\chi_{4}(t)$ initially increases with the time interval $t$ and
it reaches a peak at $t=\tau_{4}^{*}$ before decreases to zero as $t\rightarrow\infty$. 
We define the maximum correlation between A and P particles to be: $\chi_{4}^{*}=\chi_{4}(\tau_{4}^{*})$.
Both $\tau_{4}^{*}(\phi)$ and $\chi_{4}^{*}(\phi)$ are plotted in Fig.~\ref{fig:chi4}(B) and (C) respectively and shown to diverge
as $\phi\rightarrow\phi_{c}^{+}$ with critical exponents indicated in the figure. 
In particular, $\tau_{4}^{*}$ has a very similar critical exponent ($\nu_{\parallel}\simeq1.24$) to three other timescales
that we have calculated earlier in this paper. 

Note that alternatively, we may also define the dynamic susceptibility to be the fluctuations squared of $c(q,t)$:
\begin{equation}
\chi_{4}^{c}(t)=N\left\langle \Delta c^{2}(q,t)\right\rangle, \label{eq:chi4c}
\end{equation}
where $q=\frac{2\pi}{\sigma}$. 
Using this definition, we also found that the timescale for $\chi_{4}^{c}(t)$ to reach its peak value
($\tau_{4}^{c*}$) diverges with the same critical exponent as $\tau_{4}^{*}$.
However the peak height $\chi_{4}^{c*}$ does not diverge with the same critical exponent as $\chi_{4}^{*}$ due to the difference between
canonical and grand-canonical ensemble when calculating $\chi_{4}^{c}(t)$ and $\chi_{4}(t)$ (\emph{cf.} Sec.~\ref{sub:grand-canonical}).

\begin{figure} 
\begin{centering}
\includegraphics[scale=0.7]{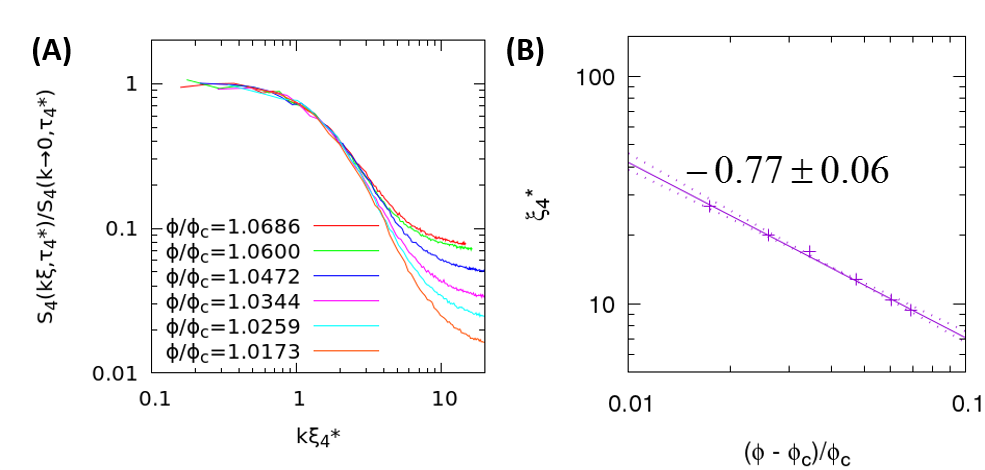}
\par\end{centering}
\caption{
\textbf{(A)} 
Dynamic structure factor $S_{4}(k,t=\tau_{4}^{*})$ as a function of the 
wavevector $k$ for different densities. 
We rescale $S_{4}(k,\tau_{4}^{*})$ as in Eq.~(\ref{eq:S4kt-scaling})
to extract the dynamic lengthscale $\xi_{4}^{*}(\phi)$. 
\textbf{(B)} 
The dynamic lengthscale $\xi_{4}^{*}(\phi)$
diverges as $\xi_{4}^{*}\sim(\phi-\phi_{c})^{-\nu_{\perp}}$ 
where $\nu_{\perp}\simeq0.77$.} 
\label{fig:S4kt4}
\end{figure}

Finally to calculate the dynamic lengthscale $\xi_{4}(t)$, we fit the dynamic structure factor $S_{4}(k,t)$ into the following scaling
form (similar to the static case in Eq.~(\ref{eq:rescaled-pair-dist-function})):
\begin{equation}
\frac{S_{4}(k,t;\phi)}{S_{4}(k\rightarrow0,t;\phi)}=F(k\xi_{4}(t;\phi)),
\quad\,\,\, k\ll\frac{2\pi}{\xi_{4}}, \label{eq:S4kt-scaling}
\end{equation}
where $F(x)$ is some universal function independent of $\phi$. Since
we are only interested in the maximum dynamic lengthscale $\xi_{4}^{*}$,
we shall fix $t=\tau_{4}^{*}$. In Fig.~\ref{fig:S4kt4}, we plot
$S_{4}(k,\tau_{4}^{*})/S_{4}(k\rightarrow0,\tau_{4}^{*})$ as a function
of $k\xi_{4}^{*}$ such that all curves corresponding to different
$\phi$'s collapse into a single function: $F(k\xi_{4}^{*})$ at small
$k$. From this plot, we extract the maximum dynamic lengthscale $\xi_{4}^{*}(\phi)$
which again diverges at the critical density with a power law: $\xi_{4}^{*}\sim(\phi-\phi_{c})^{-\nu_{\perp}}$
as $\phi\rightarrow\phi_{c}^{+}$. The critical exponent $\nu_{\perp}\simeq0.77$,
in this case, is similar to that found from the static lengthscales.
Thus close to criticality, both dynamic and static lengthscales diverge
with the same critical exponent, differing only in their prefactors.

\section{Conclusions}

\label{sec:conclusion}

In conclusion, we have introduced 
and studied numerically 
a simple and generic model of periodically 
driven suspensions which possesses a non-equilibrium absorbing phase 
transition at a critical density $\phi_c$.
In particular, our model has provided a closer connection to 
\revise{(conserved) directed 
percolation universality classes} by removing the anisotropy associated with 
sheared suspensions, which might otherwise 
affect the critical behaviour observed numerically by introducing 
a short-scale anisotropic behaviour of physical quantities~\cite{schall}.  
In this paper, both static and dynamic correlations lengths 
and associated susceptibilities are also 
measured for the first time.
In particular, the dynamic correlation length is obtained from dynamic 
heterogeneity in both space and time, in analogy to glassy 
systems~\cite{berthier-book}.
Both static and dynamic lengthscales produce similar critical exponent ($\nu_{\perp}$).
We also measured several different timescales 
independently and they are all shown to diverge \revise{again}
with a similar critical exponent ($\nu_{\parallel}$).
\revise{Our main achievement has been the demonstration that
a large number of physical observables, commonly measured 
in experimental work on colloidal suspensions, provide 
direct access to the criticality associated to the non-equilibrium 
phase transition. Thus, the mere existence of such a 
phase transition in experimental work dealing with real suspensions
subjected to an oscillatory shear flow is a question that should
be amenable to decisive experimental investigations guided by the 
present results.}

\revise{In this respect, we have established a useful analogy
with the physics of glassy systems such as intermittent dynamics
and spatially heterogeneous dynamical relaxation. The analogy 
with such materials is not only a curiosity. It also provides 
a theoretical framework and practical tools to characterize
the relevant physical observables characterizing the dynamics 
in systems close to a non-equilibrium phase transition, despite 
the distinct physical origin of dynamic correlations in both types 
of systems.} 
Our model also shows true hyperuniformity at large enough 
lengthscales, with an exponent $\lambda=1$ at criticality $\phi_c$. 
The issue of hyperuniformity 
was tackled in several recent 
studies~\cite{bartolo-hyperuniform,elsen1,levine,frenkel-hyperuniform}.

\begin{table} 
\begin{centering}
\begin{tabular}{c|ccc}
quantities & our model & DP & CDP\tabularnewline
\hline 
order parameter, $\langle f_{a}\rangle\sim(\phi-\phi_c)^\beta$ & $0.59\pm0.02$ & $0.58$ & $0.64$ \tabularnewline
order parameter decay, $f_{a}(t)\sim t^{-\alpha}$ at $\phi=\phi_c$ & $0.45$ & $0.45$ & $0.42$ \tabularnewline
fluctuations of the order parameter, $\left\langle N\Delta f_{a}^2\right\rangle_{gc}\sim(\phi-\phi_c)^{-\gamma_{gc}}$ & $0.32\pm0.02$ & $0.30$ & $0.37$\tabularnewline
timescale, $\tau\sim(\phi-\phi_c)^{-\nu_\parallel}$ & $1.26\pm0.03$ & $1.30$ & $1.23$\tabularnewline
lengthscale, $\xi\sim(\phi-\phi_c)^{-\nu_\perp}$ & $0.74\pm0.03$ & $0.73$ & $0.80$\tabularnewline
\end{tabular}
\par\end{centering}
\caption{
Critical exponents of the order parameter ($\left\langle f_{a}\right\rangle$, $f_a(t)$), 
fluctuation squared of the order parameter ($\left\langle N\Delta^{2}f_{a}\right\rangle $),
timescales ($\tau$) and lengthscales ($\xi$) as compared to directed
percolation (DP) and conserved directed percolation (CDP) \cite{lubeck-review,lubeck-book}.
\label{tab:exponent}}
\end{table}

\revise{The question of the universality class of the transition 
found in the present model is a more 
difficult problem, as the naturally expected types of 
transitions for this system, directed 
percolation and conserved directed percolation, 
are characterized by very close sets of critical exponents and 
no real qualitative difference 
between them~\cite{lubeck-review,lubeck-book}.  
Our measurements for all these critical exponents are summarised 
in Table~\ref{tab:exponent}
and for comparison, critical exponents for the directed percolation (DP) 
and conserved directed percolation (CDP) universality class are also 
shown on the table. Given how close the two sets of exponents are, it is 
no surprise that our measured exponents remain quantitatively consistent with 
both DP or CDP universality classes, despite our large numerical effort. 
We found for instance that our results are robust against finite size effects
and extensive time averaging in very large systems. 
More careful determination of critical exponents seeking to make 
a decisive statement about the DP versus CDP universality class from 
a purely computational perspective would have to involve a truly 
significant computational effort, perhaps focusing on only a handlful 
of the measurements reported in our work.

At present, it would seem that our 
critical exponents favor slightly the DP universality class, 
as can be judged by the values reported in Table~\ref{tab:exponent}.
This is a little surprising at first sight, as the main physical 
ingredient identified in lattice models to distinguish between the two families
of models is the presence of particle conservation in the microscopic
rules defining the model~\cite{lubeck-review,lubeck-book}. This 
remark forms the basis of the approach put forward
in Ref.~\cite{ramaswamy}, which suggests that the present model should 
belong to the CDP universality class. How to reconcile our numerical 
findings with this natural expectations? A first possibility 
is that a crossover to CDP values of the critical exponents 
would occur only much closer to the critical point, 
for instance because the role of particle 
conservation is only effective extremely close to the critical point.
In this view, we would interpret our measurements as being 
pre-asymptotic (despite our largest correlation length being of 
about 50 particle diameters) and, for reasons that remain to be 
elucidated, this regime appears closer to the DP behaviour.  
A second hypothesis is that our results reflect the true asymptotic 
behaviour of the model. This alternative view implies that despite 
the presence of particle conservation in the microscopic dynamics,
its effect is not felt at large enough lengthscales close to the critical 
point; in other word the critical behaviour of the model is truly 
that of directed percolation and particle conservation is 
asymptotically irrelevant.
As mentioned several times, 
we believe it will challenging to tell the two hypothesis
apart on the basis of numerical simulations only, and progress 
should come from a theoretical treatment of the critical 
behaviour of the model, which remains to be performed.}

{\color{black}
In conclusion, we hope that our simplified model will open up 
new theoretical approaches in the area of periodically driven 
non-equilibrium systems, as well as motivate novel experiments 
to determine and expose the nature of the non-equilibrium phase 
transition seen in periodically sheared suspensions.}

\ack
We thank D. Bartolo, \revise{J.-P. Bouchaud and R. Jack} 
for useful discussions. 
The research leading to these results has
received funding from the European Research Council
under the European Union’s Seventh Framework
Programme (FP7/2007-2013)/ERC Grant Agreement
No. 306845.


\section*{References} 

\begin{thebibliography}{99}

\bibitem{hinrichsen}
H. Hinrichsen, 
Nonequilibrium critical phenomena and phase transitions into absorbing states,
\emph{Adv. Phys.} \textbf{49} (2000) 815.

\bibitem{lubeck-review}
S. Lubeck, 
Universal scaling behaviour of non-equilibrium phase-transitions, 
\emph{Int. J. Mod. Phys. B} \textbf{18} (2004) 3977.

\bibitem{lubeck-book}
M. Henkel, H. Hinrichsen, and S. Lubeck,
\emph{Non-Equilibrium Phase Transitions Vol. 1. Absorbing Phase Transitions}, 
Springer, 2008.

\bibitem{Pine-Nat}
D. J. Pine, J. P. Gollub, J. F. Brady, and A. M. Leshansky,
Chaos and irreversibility in sheared suspensions,
\emph{Nature} \textbf{438} (2005) 997.

\bibitem{Pine-Nat-Phys}
L. Cort\'e, P. M. Chaikin, J. P. Gollub, and D. J. Pine,
Random organization in periodically driven systems,
\emph{Nature Physics} \textbf{4} (2008) 420.

\bibitem{CortePRL}
L. Cort\'e, S. J. Gerbode, W. Man, and D. J. Pine,
Self-Organized Criticality in Sheared Suspensions
{\it Phys. Rev. Lett.} {\bf 103} (2009) 248301.

\bibitem{bartolo-nat-comms}
R. Jeanneret and D. Bartolo,
Geometrically protected reversibility in hydrodynamic 
Loschmidt-echo experiments,
\emph{Nat. Comm.} \textbf{5} (2014) 3474.

\bibitem{ganapathy}
K. H. Nagamanasa, S. Gokhale, A. K. Sood, and R. Ganapathy,
Experimental signatures of a nonequilibrium phase 
transition governing the yielding of a soft glass,
\emph{Phys. Rev. E} \textbf{89} (2014) 062308.

\bibitem{bartolo-hyperuniform}
J. H. Weijs, R. J. Dreyfus, and D. Bartolo,
Emergent hyperuniformity in periodically-driven emulsions,
\emph{Phys. Rev. Lett.} \textbf{114} (2015) 110602.

\bibitem{filippidi}
A. Franceschini, E. Filippidi, E. Guazzelli, and D. J. Pine,
Dynamics of non-Brownian fiber suspensions under periodic shead,
\emph{Soft Matter} \textbf{10} (2014) 6722.

\bibitem{gollub}
J. S. Guasto, A. S. Ross, and J. P. Gollub,
Hydrodynamic irreversibility in particle suspensions with nonuniform strain,
\emph{Phys. Rev. E} \textbf{81} (2010) 061401.

\bibitem{cipelletti}
E. D. Knowlton, D. J. Pine, and L. Cipelletti,
A microscopic view of the yielding transition in concentrated emulsions,
\emph{Soft Matter} \textbf{10} (2014) 6931.

\bibitem{takeshi}
T. Kawasaki and L. Berthier, 
Macroscopic yielding in jammed solids is accompanied by a non-equilibrium first-order transition in particle trajectories, 
arXiv:1507.04120.

\bibitem{keim}
N. C. Keim and P. E. Arratia,
Mechanical and microscopic properties of the reversible plastic regime in a 2D jammed material,
\emph{Phys. Rev. Lett.} {\bf 112} (2014) 028302.

\bibitem{butler1}
B. Metzger and J. E. Butler,
Irreversibility and chaos: Role of long-range hydrodynamic interactions in sheared suspensions,
\emph{Phys. Rev. E} {\bf 82} (2010) 051406.

\bibitem{butler2}
B. Metzger, P. Pham, and J. E. Butler,
Irreversibility and chaos: Role of lubrication interactions in sheared suspensions,
\emph{Phys. Rev. E} {\bf 87} (2013) 052304.

\bibitem{sundararajan}
P. Sundararajan, J. D. Kirtland, D. L. Koch, and A. D. Stroock,
Impact of chaos and Brownian diffusion on irreversibility in Stokes flows,
\emph{Phys. Rev. E} \textbf{86} (2012) 046203.

\bibitem{bartolo-kurchan}
G. D\"uring, D. Bartolo, and J. Kurchan,
Irreversibility and self-organisation in hydrodynamic echo experiments, 
\emph{Phys. Rev. E} \textbf{79} (2009) 030101.

\bibitem{reichhardt3}
I. Regev, T. Lookman, and C. Reichhardt,
Onset of irreversibility and chaos in amorphous solids under periodic shear,
\emph{Phys. Rev. E} \textbf{88} (2013) 062401.

\bibitem{o-hern}
C. F. Schreck, R. S. Hoy, M. D. Shattuck, and C. S. O'Hern,
Particle-scale reversibility in athermal particulate media below jamming,
\emph{Phys. Rev. E} \textbf{88} (2013) 052205.

\bibitem{elsen1}
E. Tjhung and L. Berthier, 
Hyperuniform density fluctuations and diverging dynamic correlations in periodically driven colloidal suspensions,
\emph{Phys. Rev. Lett.} {\bf 114} (2015) 148301.

\bibitem{schmiedeberg}
L. Milz and M. Schmiedeberg,
Connecting the random organization transition and jamming 
within a unifying model system,
\emph{Phys. Rev. E} {\bf 88} (2013) 062308.

\bibitem{ramaswamy}
G. I. Menon and S. Ramaswamy,
Universality class of the reversible-irreversible transition in sheared suspensions,
\emph{Phys. Rev. E} \textbf{79} (2009) 061108.

\bibitem{reichhardt}
N. Mangan, C. Reichhardt, and C. J. Olson Reichhardt,
Reversible to irreversible flow transition in periodically driven vortices,
\emph{Phys. Rev. Lett.} \textbf{100} (2008) 187002.

\bibitem{reichhardt2}
C. Reichhardt and C. J. Olson Reichhardt,
Random organization and plastic depinning,
\emph{Phys. Rev. Lett.} \textbf{103} (2009) 168301.

\bibitem{fiocco}
D. Fiocco,
\emph{Oscillatory deformation of amorphous materials: a numerical investigation,} 
Ph. D. Thesis. Ecole Polytechnique Federale de Lausanne, 2014.

\bibitem{fiocco-foffi}
D. Fiocco, G. Foffi, and S. Sastry,
Oscillatory athermal quasistatic deformation of a model glass,
\emph{Phys. Rev. E} \textbf{88} (2013) 020301.

\bibitem{levine}
D. Hexner and D. Levine,
Hyperuniformity of critical absorbing states,
\emph{Phys. Rev. Lett.} \textbf{115} (2015) 108301.

\bibitem{frenkel-hyperuniform}
K. J. Schrenk and D. Frenkel,
Evidence for non-ergodicity in quiescent states of 
periodically sheared suspensions,
arXiv:1510.01280.

\bibitem{berthier-book}
L. Berthier, G. Biroli, J.-P. Bouchaud, L. Cipelletti, W. van Saarloos,
\emph{Dynamical heterogeneities in glasses, colloids and granular materials}, 
Oxford University Press, 2011.

\bibitem{RMP}
L. Berthier and G. Biroli, 
Theoretical perspective on the glass transition and amorphous materials,
{\it Rev. Mod. Phys.} {\bf 83} (2011) 587.

\bibitem{schall}
D. V. Denisov, M. T. Dang, B. Struth, G. H. Wegdam, and P. Schall, 
Particle response during the yielding transition of colloidal glasses, 
arXiv:1401.2106v1.

\bibitem{percier}
B. Percier, T. Divoux, and N. Taberlet, 
Insights on the local dynamics induced by thermal cycling in granular matter,
\emph{Europhys. Lett.} \textbf{104} (2013) 24001.

\bibitem{reviewPNIPAM}
P. J. Yunker, K. Chen, M. D. Gratale, M. A. Lohr, T. Still, and 
A. G. Yodh, Physics in ordered and disordered colloidal matter composed 
of poly(N-isopropylacrylamide) microgel particles, 
{\it Rep. Prog. Phys.} {\bf 77} (2014) 056601.

\bibitem{lebo}
\revise{J. L. Lebowitz, J. K. Percus, and L. Verlet,
Ensemble Dependence of Fluctuations with Application to Machine Computations,
{\it Phys. Rev.} {\bf 153}, 250 (1967).}

\bibitem{bhatia-thornton}
A. B. Bhatia and D. E. Thornton,
Structural Aspects of the Electrical Resistivity of Binary Alloys, 
\emph{Phys. Rev. B} \textbf{2} (1970) 3004.

\bibitem{stillinger-review}
S. Torquato and F. H. Stillinger,
Local density fluctuations, hyperuniformity, and order metrics,
\emph{Phys. Rev. E} \textbf{68} (2003) 041113.

\bibitem{stillinger-prl}
A. Donev, F. H. Stillinger, and S. Torquato,
Unexpected fluctuations in jammed disordered sphere packings,
\emph{Phys. Rev. Lett.} \textbf{95 }(2005) 090604. 

\bibitem{berthier-hyperuniform}
L. Berthier, P. Chaudhuri, C. Coulais, O. Dauchot, and P. Sollich,
Suppressed compressibility at large scale in jammed packings 
of size-disperse spheres,
\emph{Phys. Rev. Lett.} \textbf{106} (2011) 120601.

\bibitem{ikeda}
A. Ikeda and L. Berthier,
Thermal fluctuations, mechanical response, and 
hyperuniformity in jammed solids.
\emph{Phys. Rev. E} \textbf{92} (2015) 012309.

\bibitem{olsson-teitel}
Y. Wu, P. Olsson, and S. Teitel,
Search for hyper uniformity in mechanically stable packings 
of frictionless disks above jamming,
arXiv:1506.01948.

\bibitem{berthier-fickian}
L. Berthier, D. Chandler, and J. P. Garrahan,
Length scale for the onset of Fickian diffusion in supercooled liquids,
\emph{Europhys. Lett.} \textbf{71} (2005) 320.

\bibitem{berthier04}
L. Berthier,
Time and length scales in supercooled liquids,
{\it Phys. Rev. E} {\bf 69} (2004) 020201(R).

\end{thebibliography}
\end{document}